\documentclass[journal]{IEEEtran} 

\usepackage{graphicx} 
\usepackage{color}
\usepackage{cite}
\usepackage{graphicx}
\usepackage{amsmath} 
\usepackage{amssymb}  
\usepackage{amsfonts}
\usepackage{mathrsfs}
\usepackage{mathtools}
\usepackage{lipsum}
\usepackage[thmmarks, amsmath]{ntheorem}
\usepackage[bbgreekl]{mathbbol}
\usepackage{multirow}
\usepackage{makecell}
\usepackage{booktabs}
\usepackage{enumitem}
\usepackage{tabularx}
\usepackage{tabulary}
\usepackage{tablefootnote}

\DeclareMathOperator{\diag}{diag}

\DeclareMathOperator{\diff}{d}

\newcommand{\R}{{\mathbb R}}

\newcommand{\mc}{\mathcal}
\newcommand{\ddt}{\tfrac{\diff}{\diff \!t}}

\newcommand{\pv}{{\text{pv}}}
\newcommand{\sg}{{\text{sg}}}
\newcommand{\sm}{{\text{sm}}}
\newcommand{\tg}{{\text{tg}}}
\newcommand{\vsc}{{\text{vsc}}}
\newcommand{\pg}{{\text{pg}}}
\newcommand{\g}{{\text{g}}}

\newcommand{\dc}{{\text{dc}}}
\newcommand{\ac}{{\text{ac}}}

\newcommand{\T}{{\mathsf{T}}}

\newcommand{\loss}{{\text{loss}}}
\newcommand{\inte}{{\text{int}}}

\newtheorem{proposition}{Proposition}
\newtheorem{assumption}{Assumption}

\newcommand\blue[1]{{\color{blue}#1}}

\usepackage{xcolor}
\usepackage{cuted}

\usepackage{standalone}
\usepackage{pgf, tikz}
\usepackage{pgfplots}
\usepackage{pgfplotstable}
\usepackage{verbatim}
\usepackage{circuitikz}
\usetikzlibrary{arrows}
\usetikzlibrary{decorations.text}
\usetikzlibrary{shapes.geometric}
\usetikzlibrary{shapes.arrows}
\usetikzlibrary{arrows}
\usetikzlibrary{shapes}
\usetikzlibrary{positioning}
\usetikzlibrary{shadows}
\usetikzlibrary{patterns}
\pgfplotsset{plot coordinates/math parser=false}
\pgfplotsset{compat=newest}
\pgfplotstableset{use comma,1000 sep=\,}
\pgfkeys{/pgf/number format/.cd,fixed,precision=2} 
\usepgfplotslibrary{groupplots}

\title{
Dynamic Modeling, Analysis, and Validation of Dual-Port Grid-Forming Control for Hybrid AC/DC Systems}
\author{Irina Suboti\'c, Dominic Gro\ss{}, Alexander Winkens, Julian Jansen, Florian Klein-Helmkamp, Andreas Ulbig
\thanks{I. Suboti\'c is with ABB System Drives,
Turgi, Switzerland. D. Gro\ss{} is with the department of Electrical and Computer Engineering at the University of Wisconsin-Madison, Madison, USA. A Winkens is with Hitachi Energy, Mannheim, Germany\blue{.} J. Jansen, F. Klein-Helmkamp, and A. Ulbig are with the Institute for High Voltage Equipment and Grids, Digitalization and Energy Economics (IAEW)
RWTH Aachen University, Aachen, Germany. The corresponding author is irina.subotic@ch.abb.com.}%
}
\begin{document}
\maketitle
\begin{abstract}
This work investigates the transient and dynamical behavior of hybrid AC/DC systems using dual-port grid-forming (GFM) control. A generalized modeling framework for hybrid AC/DC networks is first introduced that accounts for converter, control, and network circuit dynamics and arbitrary network topologies. This modeling framework is applied to low-voltage networks to analyze the performance of dual-port grid-forming (GFM) control. The results demonstrate that active damping by dual-port GFM control is effective at improving the transient response and mitigating oscillations. In contrast, the steady-state response characteristics can be adjusted independently with minimal impact on damping characteristics. The dynamic model and results are validated through hardware experiments for three prototypical system architectures. Furthermore, we demonstrate that low-voltage DC distribution interfaced by AC/DC converters using dual-port GFM control, can serve both as the sole interconnection between AC distribution systems and in parallel to an AC connection, thereby enhancing the operational flexibility of low- and medium-voltage distribution networks.
\end{abstract}
\section{Introduction}
\IEEEPARstart{T}{he} power system infrastructure is experiencing significant transformations across all voltage levels that are primarily driven by the increasing integration of power electronics interfaced generation, storage, load, transmission, and distribution. This includes solar photovoltaic (PV) systems and wind turbines (WT), and battery energy storage systems (BESS), as well as the the proliferation of large direct current (DC) loads including electric vehicle (EV) charging stations, data centers, as well as high-voltage direct current (HVDC) links for long-distance energy transmission~\cite{RSV+15,SD16,GSP+21}.

At low voltage, particularly in residential areas, distributed PV generation and EV chargers are becoming increasingly common. Moreover, a substantial portion of household and residential loads already operate on DC. As a result, low-voltage direct current (LVDC) systems are emerging as a promising and cost-effective alternative to conventional AC systems, offering improved efficiency and compatibility with modern DC  technologies (see~\cite{RSV+15} and references therein).

A significant limitation to integrate renewable generation, EV chargers, and data centers, lies in insufficient transmission infrastructure, which constrains power transfer capabilities at all voltage levels. Traditionally, photovoltaic and wind generation are integrated through medium-voltage AC networks~\cite{SD16}. However, the integration of these renewable sources introduces harmonic distortion that can compromise the overall stability of the AC network. As a result, medium-voltage direct current (MVDC) network architectures are gaining attention as viable and cost-effective alternatives. MVDC systems not only alleviate transmission bottlenecks but also enable interconnection of multiple AC networks, offering enhanced flexibility and capacity in terms of both transmission and distributed generation particularly when MVDC network integrates local energy storage and renewable generation units~\cite{SD16}. While the interconnection of multiple AC systems via high-voltage DC (HVDC) links has been explored for synchronous areas~\cite{SMG+25}, similar controls can be used for MVDC and LVDC. 

Today,  converter control approaches can be broadly categorized in two major groups, AC GFM and AC GFL controls~\cite{GSP+21}. In order to achieve their objectives AC GFL controls require a stable voltage (i.e., frequency and magnitude) at the converters AC terminal and impose a stable voltage at the converter DC terminal. In contrast, AC GFM controls require a stable voltage at the converter DC terminal and impose a stable AC voltage (i.e., frequency and magnitude) at the converter AC terminal. Typically AC GFL power converters are used to, e.g., maximize the energy yield of renewables or minimize high voltage direct current (HVDC) transmission losses; but they can be also used to provide typical ancillary services (e.g., primary frequency control). However, the dynamic stability of the power system can rapidly deteriorate as the share of AC GFL resources increases~\cite{MDH+18,MBP+2019}. To resolve this issue, AC GFM are envisioned to be the cornerstone of future power systems~\cite{MBP+2019}. The prevalent AC GFM controls such as active power - frequency ($P_\ac\!-\!f$) droop control~\cite{CDA93}, virtual synchronous machine control \cite{DSF2015}, and (dispatchable) virtual oscillator control \cite{GCB+19,SG+20}, provide fast and reliable grid support~\cite{MBP+2019}. However, they may destabilize the system if the resource interfaced by the converter reach their power generation limits~\cite{TGA+20}. In contrast, machine emulation control~\cite{CBB+2015,HHH+18} retains some GFM features under power generation limits~\cite{TGA+20}. However, the literature on machine emulation control rely on a DC source that tightly controls DC voltage and does not analyze the impact of dynamics of renewables, DC networks, energy storage, and legacy synchronous generators. 

Operating hybrid AC/DC power systems using conventional AC GFM and AC GFL control generally requires assigning GFM or GFL roles to the AC and DC terminals of converters~\cite{SGP+23}. This results in significant stability challenges in hybrid AC/DC systems particularly when a fast autonomous converter response (e.g., primary frequency response) to contingencies is required~\cite{GSP+21,SGP+23,SMG+25}. In contrast, dual-port GFM control~\cite{GSP+21} can simultaneously form both AC and DC voltages and does not require GFM/GFL role assignment~\cite{SMG+25}, making it a promising solution for ensuring reliable operation in hybrid AC/DC networks. Specifically, dual-port GFM control utilizes a $v_\dc\!-\!f$ droop mechanism to simultaneously control both AC frequency and DC voltage, allowing for (i) bidirectional power support in which the direction of power flow across AC/DC interfaces autonomously adapts to the system topology, operating conditions, and available control reserves, and (ii) unified framework that generalizes conventional AC GFM and AC GFL control paradigms~\cite{SG21,GSP+21}. Previous studies~\cite{SG21,SG24} provide analytical results for small-signal stability of dual-port GFM control in hybrid AC/DC systems incorporating both legacy (e.g., synchronous generators and condensers) and modern technologies (e.g., renewables and HVDC systems). However, these works use reduced-order network models and do not provide insights on the transient response and tuning of dual-port GFM control beyond its steady-state response. At the same time, harmonic stability~\cite{ZXM+2022} and propagation of oscillations~\cite{ZZC+2022} are a significant concern and can generally only be detected when considering network-circuit dynamics that is neglected in~\cite{SG24}. A key challenge currently facing GFM deployment in hybrid AC/DC networks is the absence of validated models, analysis methods, and control tuning guidelines that facilitate a deeper understanding of the dynamics of such hybrid and heterogeneous AC/DC systems.

To this end, the first contribution of this work is a small-signal model that captures both AC and DC system dynamics including converters, resources, and network-circuit dynamics for general topologies. Subsequently, we focus on hybrid LVAC/LVDC distribution systems and analyze the transient response of the system under dual-port grid-forming (GFM) control, with particular emphasis on robustness, performance, and active damping functions. The analytical model and results are validated experimentally using a testbed comprising a synchronous generator (SG), a high-fidelity photovoltaic (PV) emulator, two voltage source converters (VSCs) with a parallel LVDC and LVAC interconnection, and (optional) connection to a public utility grid. To the best of our knowledge, this paper presents the first application and experimental demonstration of dual-port GFM control in LVDC/LVAC distribution systems and validates grid-support functions and analytical insights previously reported for high-voltage systems \cite{SG21, SG24, SG23ifac, GSP+21}.
We also examine the sensitivity of practical control implementations to measurement noise and the impact of control gains on transient performance.

This manuscript is structured as follows. An overview of the problem formulation and experimental testbed is presented in Section~\ref{sec:lab.setup}. In Sec.~\ref{sec:general.small.signal.model}, we introduce a small-signal modeling framework for hybrid AC/DC systems. Section ~\ref{sec:convinterfacedPV} through Sec.~\ref{sec:LVDCsync} investigate hybrid AC/DC system dynamics using analytical models, simulations, and hardware experiments. A synchronous generator (SG) interconnected to  photovoltaic (PV) system is investigated in Sec.~\ref{sec:convinterfacedPV}. Section~\ref{sec:LVDCasync} investigates the dynamics of a low-voltage AC system with SG connected to the utility grid through a low voltage DC connection. A low-voltage DC connection to reinforce a low voltage AC network is studied in Sec.~\ref{sec:LVDCsync}. Finally, Sec.~\ref{sec:conclusion} summarizes the key findings and outlines future research directions.
\section{Background and Problem Setup} \label{sec:lab.setup} 
This section provides a brief overview of hybrid AC/DC networks, converter control objectives, and provides and reviews results on dual-port GFM control. Moreover, we introduce the laboratory setup used to validate our theoretical results.
\subsection{Hybrid AC/DC systems}
Hybrid AC/DC networks are envisioned to encompass AC and DC networks and heterogeneous technologies ranging from legacy SGs to renewables, transmission, and storage connected by VSCs. While continental-scale hybrid AC/DC networks are emerging at high voltage levels, low voltage hybrid AC/DC networks arise from integration of renewables, storage, and DC distribution (see e.g.,~\cite[Fig.~1]{RSV+15} or~\cite[Fig.1]{SD16}).

\begin{figure}[t]
	\centering 
	\includegraphics[trim={0.65cm 0.3cm 1.05cm 0.3cm},clip, width=1\columnwidth]{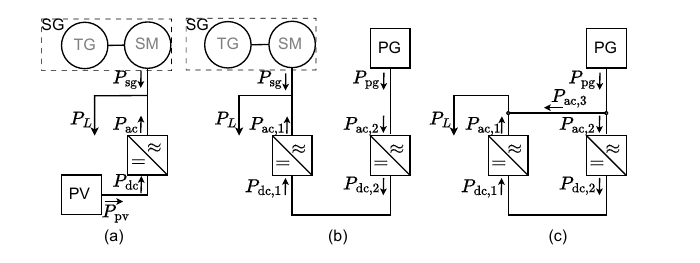} 
	\caption{Simplified illustration including SG and converter units and power-flows of the considered test cases: (a) islanded LVAC system consisting of PV and an SG, (b) LVAC system with SG connected to the utility grid via LVDC and (c) LVAC system connected to the utility grid via LVDC and LVAC.	\label{fig:simplified.setup}}
\end{figure}

In this context, dual-port GFM control can prove advantageous for three main uses in hybrid AC/DC systems (i) providing AC GFM capabilities from intermittent and variable renewable resources, (ii) interconnecting (non-synchronous) AC systems through DC networks, while providing bi-directional grid-support (e.g., inertia and fast frequency response) between the AC systems, and (iii) seamless integration of DC lines into synchronously connected AC systems. 
These use cases are reflected in the three prototypical low-voltage hybrid AC/DC networks shown in Fig.~\ref{fig:simplified.setup}. Figure~\ref{fig:simplified.setup}~(a) shows converter-interfaced generation connected to a low-voltage AC network. An islanded low voltage AC network is connected to a utility grid through an LVDC connection in Fig.~\ref{fig:simplified.setup}~(b). The capacity of a low voltage AC distribution system is increased by an LVDC connection in Fig.~\ref{fig:simplified.setup}~(c). Notably, in this work, we will also model the DC connection between renewable (e.g., PV) and their power converter as a DC network. In this context, a key question becomes how to distribute the overall control objective of stabilizing AC network frequency and voltage magnitudes and DC network voltages across legacy generation and converters interfacing AC and DC networks.

\subsection{Converter Control Objectives}
We consider three broad converter control objectives.
\begin{enumerate}
    \item \label{item:1} Dispatch: During nominal steady-state operation, the AC frequency $\omega$, voltage magnitude $V_\ac$, active $P_\ac$ and reactive $Q$ powers, DC voltage $v_\dc$, and DC power $P_\dc$ should track operating points (i.e., $\omega^\star$,$V_\ac^\star$,$P_\ac^\star$,$Q^\star$,$v_\dc^\star$,$P_\dc^\star$) prescribed by a system-level control. 
    \item \label{item:2} Unit-level stability: during disturbances (e.g., variation in the load, generation, or contingencies) the VSCs should autonomously respond and stabilize frequency, AC voltage, and DC voltage at its terminal until new a system-level control provides an updated dispatch.
    \item \label{item:3} System-level stability: Autonomous adjustment of the AC and/or DC power injections to balance energy storage elements~\cite[Sec.~III]{GSP+21} (e.g., DC link capacitors) and power flows across the hybrid AC/DC system.
\end{enumerate} 
Tracking AC active power or DC power setpoints is a common VSC control requirement. Because the VSC is a conversion element that does not produce any power the requirement to track an active power setpoints can be expressed as a requirement on internal energy balancing and requirements on terminal quantities. In this context, we need to distinguish between two main converter roles i) interfacing generation and/or storage, and ii) interconnecting AC and DC networks. For converter-interfaced renewable generation, the desired power setpoint is applied to the power source, e.g., PV (  cf.~\cite{SG23powertech}). For a converter interfacing DC and AC network, the power flow can implicitly be adjusted trough the DC voltage setpoints. In particular, for back-to-back DC connection, the desired power flow can be achieved through the DC voltage drop~\cite{SG24,SG23ifac}. By autonomously adjusting AC and/or DC power injection, internal energy storage elements are stabilized. In case of two-level converter, stabilizing a DC storage element is the same as stabilizing DC link capacitor voltage.

\subsection{Review of Converter Controls} \label{subsec:control.overview}
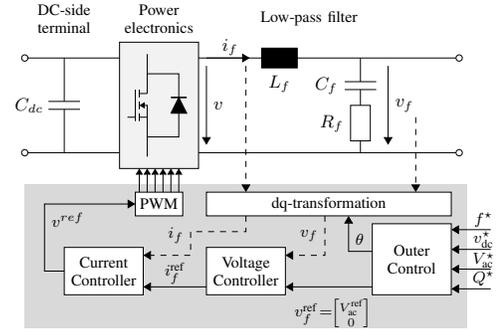
\begin{figure}[t]
    \centering
    \scalebox{.75}{\centering
\ctikzset{bipoles/thickness=0.5}
\ctikzset{tripoles/thickness=0.5}
\ctikzset{bipole label style/.style={font=\small}}
\tikzstyle{every node}=[font=\small]
\tikzstyle{every path}=[line width=0.5pt]

\begin{circuitikz}[scale=0.7]
    \draw (0.6,-1.1) -- (1.4,-1.1);
    \draw (0.6,-1.3) -- (1.4,-1.3);
    \draw (1,-1.1) -- (1,0);
    \draw (1,-1.3) -- (1,-2.4);
    \draw (2.4,0) -- (0,0) node[ocirc](){};
    \draw (2.4,-2.4) -- (0,-2.4) node[ocirc](){};
    \node[font={\footnotesize},align=center] at (1,1) {DC-side\\ terminal};
    \node[font={\small},align=center,left] at (0.65,-1.2) {$C_{dc}$};
    \draw [fill=gray!10] (2.4,0.4) rectangle (4.4,-2.8);
    \draw (2.6,-1.5) -- (2.8,-1.5);
    \draw (2.8,-0.8) -- (2.8,-1.6);
    \draw (2.9,-0.8) -- (2.9,-1);
    \draw (2.9,-1.1) -- (2.9,-1.3);
    \draw (2.9,-1.4) -- (2.9,-1.6);
    \draw (2.9,-0.9) -- (3.1,-0.9);
    \draw (2.9,-1.5) -- (3.1,-1.5);
    \draw (3.1,0.1) -- (3.1,-0.9);
    \draw[-{Triangle[scale=0.7]}] (3.1,-2.5) -- (3.1,-1.2) -- (2.9,-1.2);
    \draw (3.1,-0.4) -- (3.9,-0.4) -- (3.9,-2) -- (3.1,-2);
    \draw[fill=black] (3.7,-1.4) -- (4.1,-1.4) -- (3.9,-1) -- (3.7,-1.4);
    \draw[thick] (3.7,-1) -- (4.1,-1);
    \node[font={\footnotesize},align=center] at (3.4,1) {Power\\ electronics};
    
    \draw (4.4,0) -- (11,0) node[ocirc](){};
    \draw (4.4,-2.4) -- (11,-2.4) node[ocirc](){};
    \draw[fill=black] (6,0.25) rectangle (6.9,-0.25);
    \draw (8.5,0) -- (8.5,-0.6);
    \draw (8.5,-2.4) -- (8.5,-0.8);
    \draw[fill=white] (8.25,-1.2) rectangle (8.75,-2.1);
    \draw (8.1,-0.6) -- (8.9,-0.6);
    \draw (8.1,-0.8) -- (8.9,-0.8);
    \node[font={\small},align=center,below] at (6.45,-0.25) {$L_f$};
    \node[font={\small},align=center,left] at (8.1,-0.7) {$C_f$};
    \node[font={\small},align=center,left] at (8.25,-1.65) {$R_f$};
    \node[font={\footnotesize},align=center] at (7.2,1) {Low-pass filter};
    
    \draw[fill=gray!30, draw=none] (0,-3.2) rectangle (11.2,-6.85);
    \draw[fill=white] (2.8,-3.4) rectangle (4,-4);
    \draw[-{Triangle[scale=0.8]}] (2.9,-3.4) -- (2.9,-2.8);
    \draw[-{Triangle[scale=0.8]}] (3.1,-3.4) -- (3.1,-2.8);
    \draw[-{Triangle[scale=0.8]}] (3.3,-3.4) -- (3.3,-2.8);
    \draw[-{Triangle[scale=0.8]}] (3.5,-3.4) -- (3.5,-2.8);
    \draw[-{Triangle[scale=0.8]}] (3.7,-3.4) -- (3.7,-2.8);
    \draw[-{Triangle[scale=0.8]}] (3.9,-3.4) -- (3.9,-2.8);
    \node[font={\footnotesize},align=center] at (3.4,-3.7) {PWM};
    \draw[fill=white] (4.6,-3.4) rectangle (10.8,-4);
    \node[font={\footnotesize},align=center] at (7.7,-3.7) {dq-transformation};
    \draw[fill=white] (8.8,-4.2) rectangle (10.8,-6);
    \node[font={\footnotesize},align=center] at (9.8,-5.1) {Outer\\ Control};
    \draw[-{Triangle[scale=1]}] (8.8,-4.9) -- (8.2,-4.9) -- (8.2,-4);
    \node[font={\footnotesize},align=center,right] at (8.2,-4.6) {$\theta$};
    \draw[-{Triangle[scale=1]}] (11.8,-4.35) -- (10.8,-4.35);
    \draw[-{Triangle[scale=1]}] (11.8,-4.85) -- (10.8,-4.85);
    \draw[-{Triangle[scale=1]}] (11.8,-5.35) -- (10.8,-5.35);
    \draw[-{Triangle[scale=1]}] (11.8,-5.85) -- (10.8,-5.85);
    \node[font={\footnotesize},align=center,above] at (11.62,-4.45) {$f^\star$};
    \node[font={\footnotesize},align=center,above] at (11.62,-5) {$v_{\dc}^\star$};
    \node[font={\footnotesize},align=center,above] at (11.62,-5.5) {$V_{\ac}^\star$};
    \node[font={\footnotesize},align=center,above] at (11.62,-5.95) {$Q^\star$};
    \draw[fill=white] (4.6,-4.8) rectangle (6.6,-6);
    \node[font={\footnotesize},align=center] at (5.6,-5.4) {Voltage\\ Controller};
    \draw[dashed,-{Triangle[scale=1]}] (7.6,-4) -- (7.6,-5) -- (6.6,-5);
    \draw[-{Triangle[scale=1]}] (8.8,-5.8) -- (6.6,-5.8);
    \node[font={\footnotesize},align=center,above] at (7.8,-7) {$v_{f}^\text{ref}\!=\! \! \left [ \!\begin{smallmatrix}
        V_\text{ac}^\text{ref} \\ 0
    \end{smallmatrix} \! \right ]$};
    \node[font={\footnotesize},align=center,left] at (7.6,-4.5) {$v_{f}$};
    \draw[fill=white] (1,-4.8) rectangle (3,-6);
    \node[font={\footnotesize},align=center] at (2,-5.4) {Current\\ Controller};
    \draw[-{Triangle[scale=1]}] (4.6,-5.8) -- (3,-5.8);
    \draw[dashed,-{Triangle[scale=1]}] (5.6,-4) -- (5.6,-4.4) -- (4.2,-4.4) -- (4.2,-5) -- (3,-5);
    \node[font={\footnotesize},align=center,above] at (3.8,-5.9) {$i_f^\text{ref}$};
    \node[font={\footnotesize},align=center,left] at (4.2,-4.5) {$i_f$};
    \draw[-{Triangle[scale=1]}] (1,-5.4) -- (0.5,-5.4) -- (0.5,-3.7) -- (2.8,-3.7);
    \node[font={\footnotesize},align=center,right] at (0.5,-4.1) {$v^{ref}$};
    \draw[-{Triangle[scale=1]}] (4.6,-0.2) -- (4.6,-2.2);
    \draw[-{Triangle[scale=1]}] (9.2,-0.2) -- (9.2,-2.2);
    \draw[-{Triangle[scale=1]}] (4.6,0) -- (5.7,0);
    \node[font={\small},align=center,right] at (4.6,-1.2) {$v$};
    \node[font={\small},align=center,above] at (5.2,-0.1) {$i_{f}$};
    \node[font={\small},align=center,right] at (9.2,-1.2) {$v_{f}$};
    \draw[dashed,-{Triangle[scale=1]}] (5.6,-0.2) -- (5.6,-3.4);
    \draw[dashed,-{Triangle[scale=1]}] (9.9,-1.5) -- (9.9,-3.4);
\end{circuitikz}}
    \caption{Single-phase representation of the VSC hardware and control structure.}
    \label{fig:GFC_overview}
\end{figure}

%
%
The prevalent grid-forming control architectures for VSCs consist of outer and inner control loops (see Fig.~\ref{fig:GFC_overview}). The outer control loop provides the AC voltage waveform reference to be imposed at the converter terminal, while the inner voltage and current control track the voltage reference producing the desired ac- voltage and current waveforms at the converter terminal. Inner voltage and current controls are typically vector PI controllers with the feed-forward part to account for the impact of the (passive) RLC filter commonly used at the converter output~\cite{RL+12,PPG07}. Additionally, to reconcile the predominantly resistive lines in low voltage systems~\cite{RL+12}, an output virtual impedance (for details see~\cite{WKHU24}) is implemented to make the AC system appear predominantly inductive as seen from the converter~\cite{MM+2019}. 

Dual-port GFM control is used as outer control~\cite{SG24}. In particular, the AC voltage frequency $\omega \in \R_{>0}$ is determined by the proportional-derivative (PD) $f-v_\dc$ droop
\begin{align}\label{eq:control.law}
         \omega = \omega^\star + (k_p + {k_d s}/{(\tau_{k_d}s+1})) (v_\dc -v_\dc^\star)
\end{align}
with DC voltage $v_\dc$ and nominal frequency $\omega^\star \in \R_{>0}$. The proportional and derivative gains are $k_{p} \in \R_{>0}$ and $k_{d} \in \R_{>0}$, while $\tau_{k_d} \in \R_{\geq 0}$ is the time constant of a realizable differentiator (for $\tau_{k_d} =0$, an ideal differentiator is obtained). Notably, the derivative term vanishes in  steady-state and only influences the transient response of dual-port GFM control. Letting $s=0$ in \eqref{eq:control.law}, we obtain the steady-state response
\begin{align}\label{eq.control.steady}
\omega - \omega^\star =  k_p  (v_\dc -v_\dc^\star),
\end{align}
i.e., the gain $k_p$ fully determines the steady-state map from DC voltage and frequency and is selected to achieve the desired steady-state response. The impact of derivative gain $k_d$ and time constant $\tau_{k_d}$ is investigated in detail in Sec.~\ref{sec:convinterfacedPV} to Sec.~\ref{sec:LVDCsync}. 

The key feature of dual-port GFM control is that the propagation of power imbalances between a VSC's AC and DC terminals is controlled to induce grid support and synchronization throughout AC and DC networks. Notably, if the VSC is connected to a power source (e.g., PV or wind turbine), the power imbalance propagates to the power source which responds depending on its available headroom. In this context, \eqref{eq:control.law} can perform either approximate MPPT or provide common grid support functions depending on the operating point of renewable generation (see~\cite[Sec.~II-E]{SG24}). In the context of the VSCs connection through DC networks (e.g., LVDC, MVDC, HVDC links), \eqref{eq:control.law} propagates imbalances through the DC network. The nominal power flow across  DC networks is implicitly established through the VSC's DC voltage set points $v_\dc^\star$. Similarly, the nominal AC active power flow then directly follows from either the nominal DC power flow or the nominal operating point of the VSC power source.

Finally, the AC voltage magnitude is controlled through the low-pass filtered $Q-V$ droop~\cite{RL+12,DSF_15} 
\begin{equation}
    V_{\ac}^\text{ref}=V_{\ac}^\star+ {k_Q}/{(s\tau_Q+1)}(Q^\star-Q).
\end{equation}


\subsection{Experimental Testbed} \label{subsec:labsetup}
The laboratory setup comprises a low voltage system with AC and DC connections, two VSCs, a synchronous generator, and a connection to the utility grid.

\subsubsection{Low Voltage Grid}
\begin{figure}[t]
    \centering
    \scalebox{.85}{

\centering
\ctikzset{bipoles/thickness=0.5}
\ctikzset{tripoles/thickness=0.5}
\ctikzset{bipole label style/.style={font=\small}}
\ctikzset{inductors/scale=0.5}
\ctikzset{capacitors/scale=0.6}
\tikzstyle{every node}=[font=\small]
\tikzstyle{every path}=[line width=0.5pt]

\begin{circuitikz}[european, scale=1]

    \draw[fill=gray!30, draw=none] (-0.8,0.75) rectangle (3.1,-4.65);
    \draw[fill=gray!30, draw=none] (-0.8,-4.6) rectangle (1.15,-5.5);
\node[font={\small},align=left,anchor=west] at (-0.7,0.4) {Area AC~1};

    \draw[fill=gray!30, draw=none] (4.75,0.75) rectangle (7.5,-4.65);
    \draw[fill=gray!30, draw=none] (5.85,-4.6) rectangle (7.5,-5.5);
    \node[font={\small},align=left,anchor=west] at (4.8,0.4) { Area AC~2};

    \draw (1,0) to[sV, fill=white] (1,-0.85) -- (1,-1.1) node[ocirc](){};
    \draw (1,-1.5) -- (1,-1.8);
    \draw (1,-1.5) -- (0.75,-1.15);
    \node[font={\footnotesize},align=center] at (2.25,-0.4) {Synchronous\\ Generator};
    \node[font={\footnotesize},align=center] at (0.5,-1.3) {$S_{SG}$};
    \draw[line width=2pt] (0,-1.8) -- (2,-1.8);
    \node[font={\footnotesize},align=center] at (2,-1.6) {VK B};
    \draw (1,-1.8) to[L] (1,-3);
    \node[font={\footnotesize},align=center,rotate=0] at (0.25,-2.2) {$\text{Z}_{\text{sm}}$};
    \node[font={\footnotesize},align=center,rotate=0] at (0.25,-2.5) {$(\ell_\sm,r_{\sm})$};
    \draw[line width=2pt] (0,-3) -- (2,-3);
    \node[font={\footnotesize},align=center] at (2,-2.8) {VK 1};
    \draw (1.5,-3) to[L] (1.5,-4.5) ;  
    \node[font={\footnotesize},align=center,rotate=0] at (2.25,-3.5) {$\text{Z}_{\text{vsc}}$};
    \node[font={\footnotesize},align=center,rotate=0] at (2.25,-3.8) {$(\ell_\text{vsc},r_{\text{vsc}})$};
    \draw[line width=2pt] (0,-4.5) -- (2,-4.5);
    \node[font={\footnotesize},align=center] at (2,-4.3) {VK 5};
    \draw[line width=0.5pt,-{Triangle[scale=1.2]}] (0.5,-3) -- (0.5,-4);
    \node[font={\footnotesize},align=center,anchor=west] at (-0.8,-3.5) {Resistive\\ load};
    \draw (0.5,-4.5) to[sacdc,fill=white] (0.5,-6.5);
    \node[font={\footnotesize},align=center] at (-0.5,-5.75) {VSC 1};
    \draw (6,-0.425) node[gridnode, fill=white](){} -- (6,-1.8);
    \node[font={\footnotesize},align=center] at (7,-0.4) {Public \\ Utility \\ Grid};
    \draw[line width=2pt] (5,-1.8) -- (7,-1.8);
    \node[font={\footnotesize},align=center] at (7,-1.6) {VK A};
   \draw (6,-1.8) to[L] (6,-3.4) to[L] (6,-4.3) -- (6,-4.5);
    \node[font={\footnotesize},align=center,rotate=0] at (6.7,-2.4) {$\text{Z}_{\text{pg}}$};
    \node[font={\footnotesize},align=center,rotate=0] at (6.7,-2.7) {$(\ell_\text{pg},r_{\text{pg}})$};
    \draw[line width=2pt] (5,-4.5) -- (7,-4.5);
    \node[font={\footnotesize},align=center] at (7,-4.3) {VK 4};
    \node[font={\footnotesize},align=center] at (6.4,-3.8) {${\ell}_\text{add}$};
    \draw (6.5,-4.5) to[sacdc,fill=white] (6.5,-6.5);
    \node[font={\footnotesize},align=center] at (5.5,-5.75) {VSC 2};

    \draw (1.5,-4.5) -- (1.5,-4.8) node[ocirc](){};
    \draw (1.5,-5.2) -- (1.25,-4.85);
    \draw (1.5,-5.2) -- (1.5,-5.5) -- (5.5,-5.5) -- (5.5,-5.2);
    \draw (5.5,-4.5) -- (5.5,-4.8) node[ocirc](){};
    \draw (5.5,-5.2) -- (5.25,-4.85);
    \node[font={\footnotesize},align=center] at (1.9,-5) {$S_{AC1}$};
    \node[font={\footnotesize},align=center] at (4.9,-5) {$S_{AC2}$};
    \node[font={\small},align=center] at (3.5,-5.2) {LVAC link};

    \draw[line width=2pt] (0,-6.5) -- (2,-6.5);
    \draw (0.5,-6.5) -- (0.5,-6.8) node[ocirc](){};
    \draw (0.5,-7.2) -- (0.25,-6.85);
    \draw (0.5,-8.3) to[dcvsource, fill=white] (0.5,-7.45) -- (0.5,-7.2);
    \node[font={\footnotesize},align=left,anchor=west] at (0.9,-7.875) {PV emulator};
    \node[font={\footnotesize},align=center] at (-0.2,-7) {$S_{VDC}$};
    \draw[line width=2pt] (5,-6.5) -- (7,-6.5);
    \draw (6.5,-6.5) to[C] (6.5,-7.5) node[tlground](){};
    \node[font={\footnotesize},align=center] at (7.15,-7) {$C_{add}$};
    \draw (1.5,-7.2) -- (1.5,-7.5) -- (5.5,-7.5) -- (5.5,-7.2);
    \node[font={\small},align=center] at (3.5,-7.2) {LVDC link};
    \draw (1.5,-6.5) -- (1.5,-6.8) node[ocirc](){};
    \draw (1.5,-7.2) -- (1.25,-6.85);
    \draw (5.5,-6.5) -- (5.5,-6.8) node[ocirc](){};
    \draw (5.5,-7.2) -- (5.25,-6.85);
    \node[font={\footnotesize},align=center] at (1.9,-7) {$S_{DC1}$};
    \node[font={\footnotesize},align=center] at (4.9,-7) {$S_{DC2}$};
\end{circuitikz}}
    \caption{Schematic representation of the laboratory setup.}
    \label{fig:lab_setup_overview}
\end{figure}
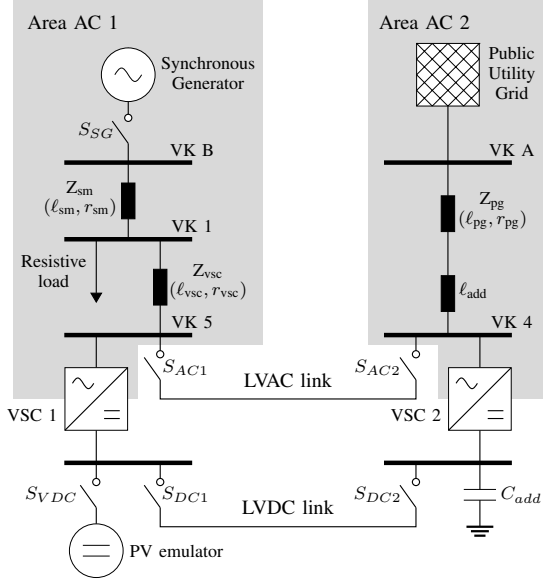

The testbed shown in Fig.~\ref{fig:lab_setup_overview} consists of two separate AC areas that can be interconnected through LVAC and/or LVDC connections. AC~2 is the public utility grid. Depending on the configuration, AC~1 can be interpreted either as a distribution feeder or low voltage microgrid.

The parameters of the LV system (i.e., cable types and lengths), are given in Table~\ref{tab:lv_setup_params}. It should be noted that area AC~2 contains an additional line inductance $\ell_\text{add}=1.5~m$H to increase the inductance to resistance ratio of the line connecting VK~4 and VK~A. Furthermore, a capacitance of $C_\text{add}=3.1~m$F is connected to the LVDC link. The additional $C_\text{add}$ is used to to dampen potential harmonic interaction between the VSCs and the DC source and suppress short-term over-voltages by slowing down DC voltage transients. Notably, the $C_\text{add}$ is equal to the DC-link capacitance of one VSC, so the total capacitance of the LVDC link is still reasonable considering the voltage level and power rating of the VSCs.

\begin{table}[t]
    \centering
    \caption{Laboratory setup parameters}
    \begin{tabular}{l l l l l}
        \multicolumn{3}{l}{Nominal operational parameters}                                                     \\ \hline \hline
        \multicolumn{2}{l}{AC voltage (phase-to-phase) and frequency}     & $400\,$V, $50\,$ Hz    \\                   
        \multicolumn{2}{l}{DC voltage range}            & $650\,$ V$\, -\, 850\,$V          \\ \hline
        \\
        \multicolumn{2}{l}{ Cable parameters} \\ \hline \hline                                                                       
        Line (see Fig~\ref{fig:lab_setup_overview})                   & Type        &   Length [m]                     \\ \hline
        L$_\sm$                    & NAYY 4x240          & $4$                    \\ 
        L$_\vsc$                  & NAYY 4x35         & $25$                  \\ \hline
        \multirow{2}{*}{L$_\pg$} & NAYY 4x240        & $5$                      \\
                                & NAYY 4x35         & $215$             \\ \hline
        LVAC                    & NAYY 4x150        & $119$         \\
        LVDC                    & H07RN-F 2x6       & $40$         \\ \hline
    \end{tabular}
    \label{tab:lv_setup_params}
\end{table}

\subsubsection{Voltage Source Converters (VSCs)} \label{subsec:LabGFM}
The test setup features two identical voltage source converters programmed through code generation from MATLAB/Simulink. A single-phase representation of the converter hardware and control architecture is shown in Fig.~\ref{fig:GFC_overview}. Each converter consists of a DC-link capacitor, a three-phase two-level switching stage, and an AC-side LC low-pass filter. Table~\ref{tab:GFC_hardware_params} summarizes the relevant VSC specifications and current and voltage PI control gains used in the testbed with per unit base given by the rated VSC voltage and power. The energy stored in the DC link capacitor is equivalent to approximately $60~m$s of rated power. While $5-10~m$s suffice for some GFL applications, providing GFM  functions with two-level VSCs may require approximately $40-70~m$s of energy storage. Finally, it should be noted that the converters also comprise permanently connected discharging resistors at the DC-side for safety reasons. However, the discharge time constant of the LVDC-link is approx. $146$~s, which is negligible compared to the transient time scales considered in this work.

\begin{table}[t]
    \centering
    \caption{Specifications of the converters}
    \begin{tabular}{l l}
         \multicolumn{2}{l}{Converter Hardware} \\ \hline \hline
         Nominal power                          & $22\, k$VA                      \\
         Nominal voltage & 800\,V \\
         Filter parameters                      & $L_f=1.18\, m$H, $C_f=20\,\mu$F, $R_f=0.5\,\Omega$\\
         Switching freq.                    & $f_{sw}=20\, k$Hz \\
         DC capacitance                         & $C_{dc}=3.1\,m$F \\ \hline  
         \multicolumn{2}{l}{ } \\
         \multicolumn{2}{l}{Controller parameters} \\ \hline \hline
         QV-droop                    & $k_Q=0.05$~p.u., $\ \tau_Q=0.2\,m$s \\
         Inner voltage loop      & $k_{pv}=0.9$~p.u.,  $ \ \ k_{iv}=1\, $p.u., \ \  $f_{\text{bw}}\approx 283$\,Hz  \\
     Inner current loop      & $k_{pc}=1$~p.u., $ \ \ \ \ \ k_{ic}=0.4$~p.u., $f_{\text{bw}}\approx 567$\,Hz \\
     Virtual impedance & $\tilde{\ell} \coloneqq 2.3\,m$H \\ \hline
    \end{tabular}
    \label{tab:GFC_hardware_params}
\end{table}


\subsubsection{Synchronous Generator (SG)}
The SG emulates the dynamics of small-scale synchronous machines connected to LV grids. The rotor of the SG is driven by an induction machine that is controlled to emulate the dynamics of a turbine/governor system. The controller is split into a voltage controller (AVR) and a speed droop governor (see \cite[Fig~2~(a), (c)]{Winkens.2022}. Table~\ref{tab:sg_params} summarizes the main specifications of the SG and its controller parameters. Further details on the SG can be found in~\cite{Erlinghagen.2017,Winkens.2022}.


\begin{table}[t]
    \centering
    \caption{Specifications of the synchronous generator}
    \begin{tabular}{l l}
		\multicolumn{2}{l}{Synchronous generator parameters} \\ \hline \hline
		Nominal power					& $S_{n,\sg}=105\,k$VA						\\
		Maximum active power$^\dagger$			& $P_{\max,\text{sg}}=50\, k$W						\\
		Nominal AC voltage (phase-to-phase)	& $V_{n,\text{sg}}=400\,$V							\\
		Nominal rotor speed					& $n_r=25\,\mathrm{s}^{-1}$					    \\
		Total inertia of the test bench     & $H=0.1417\,$s                				\\
		Droop and damping coefficients (governor)	    & $k_\tg=20$~p.u., $k_\omega=0.5$~p.u.\\			           
		Lag elements (governor)				& $T_1=30\,ms,\,T_2=100\,m$s				\\ \hline
        \multicolumn{2}{l}{$^\dagger$\footnotesize{ base power for the per-unit values in this table.}} \\
    \end{tabular}
    \label{tab:sg_params}
\end{table}

\subsubsection{PV Emulator}

The DC source of VSC~1 has a rated power of $30$~kW and is able to provide bidirectional power flows. For the experiments presented in this paper, it is either operating as PV emulator to examine converter-interfaced renewable generation, or disconnected. In the PV emulation mode, the output voltage follows a fixed voltage-current-characteristic. The underlying model is implemented according to the test procedure for measuring the efficiency of MPPT-tracking in PV inverters outlined in the standard DIN EN 50530. The cSi-technology parameter set is chosen \cite{DINEN50530.2013}. Considering the specifications of the VSCs, the maximum power point is emulated at $650\,$V and $28\,$A, corresponding to $18.2\,$kW. The open circuit voltage and short circuit current resulting from the model are $812.5\,$V and 31.1\,A. In the experimental tests, we use $V_{dc}=740\,$V to be able to observe possible DC voltage fluctuations. This corresponds to a curtailment of $28\%$ (i.e., slightly higher than typical curtailment levels). Finally, we emphasize that, using dual-port GFM control, the sensitivity $k_\pv = 3.4581$~p.u. of the $v_{\dc}-P_{\dc}$ characteristic determines the VSC $P_\ac-f$ droop response (see~\cite[Fig.~4]{SG24}). 


\subsubsection{Testbed configurations and topologies}
By opening and closing switches the hardware setup depicted in Fig.~\ref{fig:lab_setup_overview} can be configured to match the configurations illustrated in Fig.~\ref{fig:simplified.setup}. If AC~1 operates in isolation as in Fig.~\ref{fig:simplified.setup}~(a), a VSC and/or SG is feeding the load. Hence, this allows for the representation of an islanded LV system or microgrid. By connecting the the PV emulator, this configuration allows to validate dual-port GFM control for PV  generation. In contrast, area AC~2 is connected to the utility grid. Connecting AC~1 and AC~2 solely through the VSCs and LVDC link results in the configuration shown in Fig.~\ref{fig:simplified.setup}~(b). Notably, in this setting the utility grid exhibits stiff frequency dynamics resembling an infinite bus. This configuration can be used to test the challenging case of interconnecting two AC areas with vastly different frequency dynamics through a DC connection. Finally, connecting the resistive load in AC~1 to AC~2 with both LVDC and LVAC links, as in Fig.~\ref{fig:simplified.setup}~(c), allows to test reinforcing a low-/medium-voltage grid using an LVDC connection.



\section{Dynamic model}\label{sec:general.small.signal.model}
In this section we present transfer functions of the individual component models, derive dynamical models of AC and DC networks, obtaining an overall hybrid AC/DC system model.

\subsection{Unit models}\label{subsec:device.models}
The transfer functions $G_\sm$, $G_\vsc$, $G_\pv$ of the conversion elements (i.e., SM and VSC) and PV are obtained by applying Laplace transform to linearized unit models (cf.~\cite[Sec.~II~C-D]{SG24}) while the transfer function $G_\tg$ of the turbine/governor system directly follows from \cite[Fig.~2~(c)]{Winkens.2022}. Thus, we have
\begin{align*}
    G_\sm&\coloneqq G_{\omega,P}= 1/(J\omega_r^\star s),\\
    G_\vsc&\coloneqq G_{v_\dc,P}=1/(C_\dc v_\dc^\star s),\\
    G_\pv&\coloneqq G_{v_\dc,P}=-k_\pv,\\
     G_\tg &\coloneqq G_{\omega,P}=-(k_\omega + k_\tg G_1(s) G_2(s)) S_{n,\sg}/P_{\max,\sg} 
\end{align*}
with $J \coloneqq 2HS_{n,\sg}/{\omega_r^\star}^2$, $\omega_r^\star = 2\pi n_r$ and $G_1(s)\coloneqq1/(T_1s+1)$ and $G_2(s)\coloneqq1/(T_2s+1)$. Finally, the transfer function of the dual-port GFM control follows from~\eqref{eq:control.law}, i.e.,
$ G_\text{ctr}\coloneqq G_{v_\dc,\omega}(s) = k_p +k_ds/(\tau_{k_d}s+1)$. Here, the notation $G_{u,y}(s) \coloneqq y(s)/u(s)$ refers to transfer function from input $u(s)$ to output $y(s)$. With a slight abuse of notation, we interchangeably use, e.g., $G_\sm$ and $G_{\omega,P}$.

\subsection{Network models}
To model the hybrid AC/DC system, we derive small-signal models for the AC and DC networks.
\subsubsection{AC network} Wirthout loss of generality, we lump (steady-state) virtual impedance of each VSC node into the AC network model. Considering the virtual inductances $\tilde{\ell}_n,\tilde{\ell}_k \in \R_{\geq 0}$ and resistances $\tilde{r}_n,\tilde{r}_k \in \R_{\geq 0}$  of the nodes $n$ and $k$ we define $\tilde{\ell}_{nk}\coloneqq \tilde{\ell}_n+\tilde{\ell}_k$, $\tilde{r}_{nk}\coloneqq \tilde{r}_n+\tilde{r}_k$. If node $n$ corresponds to a node without virtual impedance (i.e., load node, SM, or VSC without virtual impedance), then $\tilde{r}_n = \tilde{\ell}_k =0$. Then, the dynamics of the current $i_{nk}\in \R^2$ (in $dq$ frame) flowing through the connection between nodes $n$ and $k$ are given by 
\begin{align*}
    \ell_{nk} \ddt i_{nk}\! = \!-((r_{nk}+\tilde{r}_{nk})\!+\!j\omega^\star (\ell_{nk}+\tilde{\ell}_{nk}))I_2i_{nk} +v_n \!-v_k
\end{align*}
where $v_{n}\in \R^2$, and $v_{k}\in \R^2$ are the node voltages and $\ell_{nk} \in \mathbb{R}_{>0}$ and $r_{nk} \in \mathbb{R}_{>0}$ denote the line inductance and resistance. Defining $\theta_{nk}\coloneqq\theta_n-\theta_k$ (with $\theta_{nk}=-\theta_{kn}$) and using the same steps as in~\cite[Sec.~II-B]{G22}, we linearize the active power $p_{\ac,kn}=p_{\ac,nk}=v^\text{T}_{n}i_{nk}$ around the trivial solution to obtain
\begin{align*}
    p_{\ac,nk}(s) = \frac{k_{nk}{V_\ac^\star}^2\omega^\star}{\ell_{nk}(\underbrace{s^2+2\tilde{\rho}_{nk}s+\tilde{\rho}_{nk}^2 +(k_{nk}{\omega^\star})^2)}_{\eqqcolon g_{\theta,p,nk}(s)}} \theta_{nk}(s),
\end{align*}
where $k_{nk}\coloneqq (\tilde{\ell}_{nk}+\ell_{nk})/\ell_{nk}$ and $\tilde{\rho}_{nk} \coloneqq (\tilde{r}_{nk}+r_{nk})\ell_{nk}$.

\subsubsection{DC network} If nodes $n$ and $k$ with DC voltages $v_{\dc,n} \in \R_{\geq 0}$, $v_{\dc,k} \in \R_{\geq 0}$ are connected by a DC link, the dynamics of the current are given by $\ell_{dc,nk} \ddt i_{\dc,nk} = -r_{\dc,nk}i_{\dc,nk} +v_{\dc,n}-v_{\dc,k}$ and $p_{\dc,nk}=v_{\dc,n}i_{\dc,nk}$. Defining $v_{\dc,nk} \coloneqq v_{\dc,n}-v_{\dc,k}$ and $g_{v_\dc,p,nk}(s) \coloneqq \ell_{\dc,nk}s+r_{\dc,nk}$, and linearizing at $v_{\dc,n}^\star  \in \R_{\geq 0}$ and $v_{\dc,k}^\star \in \R_{\geq 0}$ results in the power flow across the DC link and DC link losses
\begin{align*}
    p_{\dc,nk}(s)\! &=\! v_{\dc,n}^\star v_{\dc,nk}(s)/g_{v_\dc,p,nk}(s), \\
    p_{\dc,\text{loss},nk}(s) \!&= \!v_{\dc,nk}^\star  v_{\dc,n}(s)/\g_{v_\dc,p,nk}(s).
\end{align*}
The DC network dynamics are linearized around arbitrary power flow, i.e., $v^\star_{\dc,n}=v^\star_{\dc,k}$ does not need to hold but $g_{v_\dc,p,nk}(s) = g_{v_\dc,p,kn}(s)$ and $v_{\dc,nk}=-v_{\dc,kn}$ hold. 
If $v^\star_{\dc,n}= v^\star_{\dc,k}$ holds, then $p_{\dc,nk}(s)=-p_{\dc,kn}(s)$ and $p_{\dc,\text{loss},nk}(s)=p_{\dc,\text{loss},kn}(s)=0$. 

\subsection{Model of the overall hybrid system}
To model the overall interconnected hybrid AC/DC network, we use the graph-based modeling approach described in detail in~\cite[Sec.~II~B]{SG24}. The set of AC nodes (i.e., SMs, VSCs, load nodes) is denoted by $\mc N_\ac$  and $\mc N_\dc$ denotes the DC node set (i.e., VSC DC terminals and interior DC nodes without converters). Note that the VSC nodes are in $\mc N_\ac \cap \mc N_\dc$. The AC and DC edge sets $\mc E_\ac$ and $\mc E_\dc$ and model AC and DC connections. We assume that the overall hybrid AC/DC network and its graph are connected. A crucial difference to~\cite{SG24} is that this work considers the dynamics of the power flow across the edges.

For all $n\in\mc N_\ac$, the power injection is given by $p_{\ac,n}(s) = \sum_{k,(n,k)\in \mc E_\ac} p_{\ac,nk}(s)$. Notably, if $n\in\mc N_\ac$ is a load node (i.e., not SM or VSC nodes), then $p_{\ac,n}(s)$ is equal to the load power $P_{L,n}(s)$. Denoting the oriented AC incidence matrix by $B_\ac \in \{-1,0,1\} \in \R^{|\mc N_\ac|\times |\mc N_\ac|}$, we have $[p_{\ac,1}(s) \ \ldots \ p_{\ac,|\mc N_\ac|}(s)]^\text{T} = B_\ac \diag\{ k_{nk}{V_\ac^\star}^2\omega^\star/(\ell_{nk}g_{\theta,p,nk}(s)) \} B_\ac^\text{T} [\theta_{1}(s) \ \ldots \ \theta_{|\mc N_\ac|}(s)]^\text{T}$.

Similarly, for $n \in \mc N_\dc$, $p_{\dc,n}(s) = \sum_{k,(n,k)\in \mc E_\dc} p_{\dc,nk}(s)$ and $p_{\dc,\text{loss},n}(s) = \sum_{k,(n,k)\in \mc E_\dc} p_{\dc,\text{loss},nk}(s)$. Using $B_\dc \in \{-1,0,1\} \in \R^{|\mc N_\dc|\times |\mc N_\dc|}$ to denote the oriented DC incidence matrix, we have $[p_{\dc,1}(s) \ \ldots \ p_{\dc,|\mc N_\dc|}(s)]^\text{T} = B_\dc \diag\{v_{\dc,n}^\star/(g_{v_\dc,p,nk}(s)) \} B_\dc^\text{T} [v_{\dc,1}(s) \ \ldots \ v_{\dc,|\mc N_\dc|}(s)]^\text{T}$ and $[p_{\dc,\loss,1}(s) \ \ldots \ p_{\dc,\loss,|\mc N_\dc|}(s)]^\text{T} = \diag\{p_{\dc,\loss,n}\}$.

Without loss of generality, we order the AC nodes such that the last $n_{\inte}$ nodes correspond to the load nodes, i.e., $P_L(s) = [p_{\ac,|\mc N_\ac|-n_\inte+1}(s) \ldots p_{\ac,|\mc N_\ac|}(s)]^\T$ and the first $|\mc N_\ac|-n_\inte$ nodes correspond to SM and VSC nodes $p_{\bar{\ac}}(s) = [p_{\ac,1}(s) \ldots p_{\ac,|\mc N_\ac|-n_\inte}(s)]^\T$. Moreover, we define $\theta_{\bar{\ac}}(s) = [\theta_{1}(s) \ldots \theta_{|\mc N_\ac|-n_\inte}(s)]^\T$, $\theta_L(s) = [\theta_{|\mc N_\ac|-n_\inte+1}(s) \ldots \theta_{|\mc N_\ac|}(s)]^\T$. Finally, we can write the AC network model as
\begin{align}\label{eq:acnetL}
    \begin{bmatrix}
        p_{\bar{\ac}}(s) \\
        P_{L}(s)
    \end{bmatrix} = \begin{bmatrix}
        L_{\bar{\ac}}(s) & L_{{\bar{\ac}},L}(s) \\
        L_{L,{\bar{\ac}}}(s) & L_{L}(s)
    \end{bmatrix}   \begin{bmatrix}
       \theta_{\bar{\ac}}(s) \\
        \theta_{L}(s)
    \end{bmatrix}.
\end{align}
We emphasize that  $\theta_{\bar{\ac}}(s)$ and $P_{L}(s)$ are independent variables and $p_{\bar{\ac}}(s)$ and $\theta_{L}(s)$ are dependent variables. Therefore, a generalization of Kron-reduction~\cite{DB2013} is used to remove $\theta_{L}(s)$ from the model and map $P_{L}(s)$ to SM and VSC nodes. This results in $\theta_L = L_L^{-1}(P_L-L_{L,\bar{\ac}}\theta_L)$ and the reduced-order model 
\begin{align} \label{eq:refuced.order.pf}
    p_{\bar{\ac}}(s)=  G_{\bar{\ac}}(s)\theta_{\bar{ac}}(s) + G_L(s)P_L(s)
\end{align}
with $G_{\bar{\ac}}(s)\coloneqq L_{\bar{\ac}}(s)-L_{\bar{\ac},L}(s)L_{L}^{-1}(s)L_{L,\bar{\ac}}(s)$ and $G_L(s)\coloneqq L_{\bar{\ac},L}(s)L_L^{-1}(s)$. The model \eqref{eq:refuced.order.pf} immediately raises the question of when $L_L(s)$ is invertible.
\begin{proposition}
    $L_L(s)$ is invertible for almost all $s\in \mathbb{C}$.
\end{proposition}
\begin{IEEEproof}
        For all $s\in \mathbb{R}_{\geq 0}$, $L_L(s)$ is loopy Laplacian and hence, always invertible~\cite{DB2013}. Moreover, $L_{L}(s)$ is a matrix of rational polynomials. Therefore, all $s\in \mathbb{C}$ for which $L_{L}(s)$ is not invertible are values for which the determinant of $L_L(s)$ is zero. Consequently, $s \in \mathbb{C}$ for which $L_L(s)$ is not invertible is a zero-measure set. 
\end{IEEEproof}
We note that any $s\in \mathbb{C}$ for which $L_L(s)$ is not invertible corresponds to poles of $L_L^{-1}(s)$ and consequently, poles of the transfer functions $G_{\bar{\ac}}(s)$ and $G_L(s)$. In other words, any $s\in \mathbb{C}$ for which $L_L(s)$ is not invertible is a pole of the system \eqref{eq:refuced.order.pf}. Hence, to guarantee stability of \eqref{eq:refuced.order.pf}, $L_L^{-1}(s)$ has to be stable. Consequently, we require the following assumption. 
\begin{assumption}\textbf{(Stability requirement of the reduced-order AC network \eqref{eq:refuced.order.pf}} \label{assump:invert.stable}
    Inverse of $L_{L}(s)$ is stable.
\end{assumption}
Assumption~\ref{assump:invert.stable} is trivially satisfied if i) The inductive-resistive ratio of the AC lines is the same within a connected AC area and ii) if there is only one interior node between any two conversion nodes (i.e., SM or VSC). While analytically checking Assumption~\ref{assump:invert.stable} for arbitrary line parameters and network structures is not a trivial task, Assumption~\ref{assump:invert.stable} can easily be checked numerically for a given system.

\begin{figure}[t]
	\centering 
    	\includegraphics[trim={1.5cm 0.2cm 1.25cm 0.175cm}, clip, width=0.8\columnwidth]{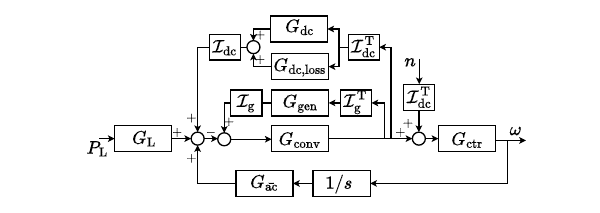} 
	\caption{Block diagram of the hybrid AC/DC small-signal model. Nominal power flows are flowing out of the units, as illustrated in~\cite[Fig.~3]{SG24}.\label{fig:blk.diag_general}}
\end{figure}

To construct an interconnected small-signal model of the hybrid AC/DC system, we introduce the diagonal matrix $G_{\text{conv}}$ of transfer functions of energy conversion (i.e., SMs and VSCs). Moreover, $G_\text{gen}$ is a diagonal matrix of transfer functions of generation resources (i.e., PV, turbine/governor). To model the interconnection of generation and conversion, we introduce the interconnection matrix $\mc I_\g$ whose entry $\{\mc I_\g\}_{i,j}$ is one if the $j$-th resource is connected to the $i$-th energy conversion unit and zero otherwise. Similarly, $\mc I_\dc$  models the interconnection of energy conversion units to the DC network, i.e., the entry $\{\mc I_\dc\}_{i,j}$ is one if the $i$-th AC node is a VSC connected to DC node $j$ and zero otherwise. Finally, $G_\text{ctr}$ is a diagonal matrix of VSC control transfer functions. If element $i$ corresponds to a VSC, $G_{\text{ctr},i}$ is given by corresponding converter control transfer function. In contrast, if an element $i$ corresponds to a SM, then $G_{\text{ctr},i}=1$. The small-signal model of the hybrid AC/DC system is illustrated in Fig.~\ref{fig:blk.diag_general}. 

\section{Islanded low voltage system} \label{sec:convinterfacedPV}
In this section, we investigate the dynamics of AC~1 operating islanded with PV and SG (see Fig.~\ref{fig:simplified.setup}~(a)). We use the small-signal model, developed in Sec.~\ref{sec:general.small.signal.model} to analyze the impact of dual-port GFM control gains and validate the results in hardware experiments. For clarity of the presentation, the PV DC voltage and power base are chosen as their MPP values.

\subsection{Small-signal model}\label{sec:PV.small.signal.model}
\begin{figure}[b]
	\centering 
    	\includegraphics[trim={0.75cm 0.3cm 1.25cm 0.425cm},clip, scale=0.97]{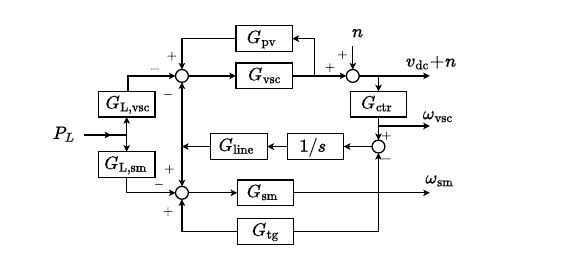} 
	\caption{Block diagram of the test case depicted in Fig.~\ref{fig:simplified.setup}~(a).	\label{fig:blk.diag1}}
\end{figure}

With $G_\pv$, $G_\vsc$, $G_\sm$, $G_\tg$  introduced in Sec.~\ref{subsec:device.models}, the small-signal model of the system (see Fig.~\ref{fig:simplified.setup}~(a)) is shown in Fig.~\ref{fig:blk.diag1}. 
Using $k_\ell \coloneqq (\tilde{\ell}+\ell_\vsc)/\ell_\vsc$, ,  $\mathbb{g}_\sm(s)\coloneqq s^2+2\rho_\sm s +\rho_\sm^2+{\omega^\star}^2$, $\rho_\sm = r_\sm/\ell_\sm$, and $\mathbb{g}_\vsc(s)\coloneqq s^2+2\rho_\vsc s +\rho_\vsc^2+(k_{\ell}\omega^\star)^2$, and $\rho_\vsc \coloneqq r_\vsc/\ell_\vsc$, the AC line dynamics become
\begin{align*}
    G_\text{line} \! &= \!{V_\ac^\star}^2\omega^\star k_l /\left(\ell_{\vsc}\mathbb{g}_\vsc(s)+\ell_{\sm}k_l\mathbb{g}_\sm(s)\right),\\
    G_{L,\sm} \!&= \! \ell_\vsc \mathbb{g}_\vsc(s) /\left(\ell_{\vsc}\mathbb{g}_\vsc(s)+\ell_{\sm}k_l\mathbb{g}_\sm(s)\right),\\
    G_{L,\vsc} \!&=\! k_l\ell_\sm \mathbb{g}_\sm(s) /\left(\ell_{\vsc}\mathbb{g}_\vsc(s)+\ell_{\sm}k_l\mathbb{g}_\sm(s)\right).
\end{align*}
Assumption~\ref{assump:invert.stable} is trivially satisfied because all coefficients of $\mathbb{g}_\vsc(s)$ and $\mathbb{g}_\sm(s)$ are positive.

\subsection{Baseline VSC control tuning}
We first use analytic steady-state and stability results to obtain an initial set of control gains.
\subsubsection{Steady-state}
Assuming a lossless converter (i.e., $P_{ac}=P_{dc}$) the steady-state frequency is $ \omega_\sg = \omega_\vsc = -  ( \kappa_\tg^{-1} + \kappa_\pv^{-1})^{-1}P_L$ with (effective) droop coefficients $\kappa_\tg \coloneqq k_\tg^{-1}$ and $\kappa_\pv= k_p/k_{\pv}$ (also see \cite[Prop.~3]{SG24}). Using $ \omega_\vsc = k_p  v_\dc$ we have $ v_\dc = -  (k_p( \kappa_\tg^{-1} + \kappa_\pv^{-1}))^{-1} P_L$. Moreover, the steady-state change in power provided by the SG and PV is given by $ P_\tg = \kappa_\tg^{-1}  ( \kappa_\tg^{-1} + \kappa_\pv^{-1})^{-1} P_L$ and $ P_\pv= \kappa_\pv^{-1}  ( \kappa_\tg^{-1} + \kappa_\pv^{-1})^{-1} P_L$. In other words, both SG and PV respond to disturbances, with contribution inversely proportional to their (effective) droop coefficients expected to be prescribed by grid codes, system operators, or aggregators. For the purpose of this study, we select $k_p \in \{0.05,0.025\}$ such that, in steady-state, a $5$\% and $10$\%  VSC DC bus voltage deviation maps to a frequency deviation of $0.25$\% (i.e., $125$~mHz in a $50$~Hz system). In the SG base, this results in $k_{\pv}/k_{p} = 50.35$ for $k_p=0.025$ and $k_{\pv}/k_{p} = 25.17$ for $k_p=0.05$. In other words, the PV unit provides a stiffer response than the SG with $k_{\tg}=20$.

Finally, this configuration of the testbed can also be interpreted as an aggregate AC system (SG) and aggregate DC (micro)grid (DC source) interconnected by a VSC. In this case, the proportional VSC gain $k_p$ can be used to adjust the power sharing between the two networks.


\subsubsection{Stability}
To obtain a range for the derivative gain $k_d$, we use analytical stability conditions that neglect network circuit dynamics~\cite[Thm.~1]{SG24}. 
Notably, the stability conditions~\cite[Cond.~1]{SG24} requires at least one unit provides steady-state droop (i.e., the SG or curtailed PV). Moreover, the topology of AC~1 satisfies~ \cite[Cor.~2~\textit{iii}]{SG24}. It remains to show that ~\cite[Cond.~2]{SG24} holds. Recall that the (emulated) PV is curtailed to demonstrate GFM functions with frequency response. In that case, \cite[Cond.~2]{SG24}  requires that the VSC control satisfy $k_{d}<4C_{dc}k_{p}/k_\pv$, where $k_\pv$ is a linearization of the PV $P$-$v_\text{dc}$ characteristic around the nominal operating point~\cite[Fig.~4]{SG24}. For the parameters of the testbed, this results in the requirement $k_d< 9.9040\, k_p$. To meet the stability condition, we select, in per unit, $(k_p,k_d)\in\{(0.025,0.01),(0.025,0.005),(0.05,0.01)\}$.

\subsection{Analysis of the dynamic response}
Next, we use the small-signal model to analyze the impact of the control gains.  
\subsubsection{Realizable-differentiator time constant $\tau_{k_d}$} 
\begin{figure}[b]
	\centering 
	\includegraphics[scale=0.63]{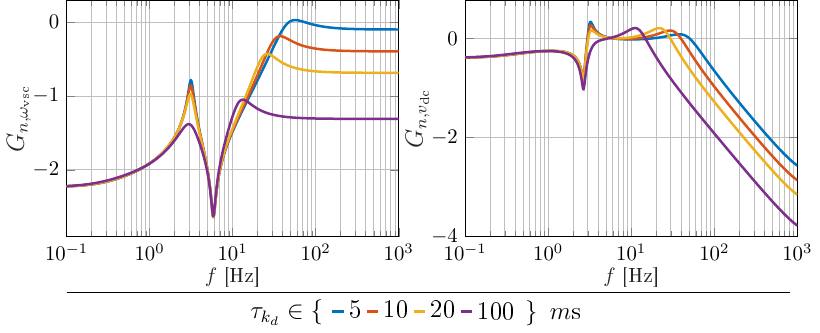} 
	\caption{Bode magnitude plot of the transfer function from measurement noise $n$ to frequency $\omega_\vsc$ (left) and DC voltage $v_\dc$ (right) and differentiator time constants $\tau_{k_d} \in \{ 5,10,20,100\}$~ms.	\label{fig:test3.tau.kd}}
\end{figure}

A crucial aspect is to differentiate between the impact of $\tau_{k_d}$ on (i) the sensitivity of the VSC to DC voltage measurement noise, and (ii) the response of the DC voltage to the load $P_L$. To this end, Fig.~\ref{fig:test3.tau.kd} illustrates the impact of the time
constant $\tau_{k_d}$ on the gain of the transfer function from DC voltage measurement noise $n$ to the VSC's frequency, $G_{n,\omega_{\vsc}}$ (left) and DC voltage, $G_{n,v_{\dc}}$ (right). As expected from the controller transfer function \eqref{eq:control.law}, reducing $\tau_{k_d}$ increases the high-frequency gain of the transfer function from measurement noise to VSC frequency and DC voltage.
Notably, magnitude of $G_{n,v_{\dc}}$ rolls off as a consequence of $v_\dc$ being the integral of the active power imbalance (cf.~\cite[(3)]{SG24}). Moreover, we observe that increasing the time constants  $\tau_{k_d}$ provides increased damping in the high frequency range. In contrast, the damping in the medium frequency range only changes significantly for large time constants $\tau_{k_d}$. Similar features can be observed for $G_{n,\omega_{\vsc}}$. Finally, we note that the same conclusions hold for $G_{P_L,\omega_{\vsc}}$. On the other hand, decreasing $\tau_{k_d}$ results in increased damping of the load power to DC voltage dynamics $G_{P_L,v_{\dc}}$ in the range of approx. $5$~Hz to $100$~Hz. For reasons of space, the corresponding Bode plots are omitted. Consequently, the choice of the derivative time constant is a trade-off between its impact on the medium frequency dynamics, damping of the response to load changes, and sensitivity to noise in the high frequency range. Overall, we can conclude that the appropriate choice of $\tau_{k_d}$ won't have significant impact on the dynamics.

\subsubsection{Impact of $k_p$ and $k_d$}
\begin{figure}[b]
	\centering 
	\includegraphics[width=\columnwidth]{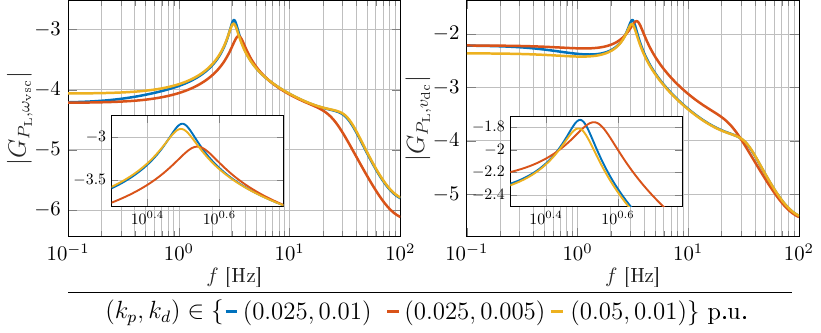} 
	\caption{Bode magnitude plot of the transfer functions from the load $\Delta P_L$ to the VSC frequency $\omega_\vsc$ (left) and DC voltage $v_\dc$ (right) for the control gains, $(kp,kd)\in\{(0.025,0.01),(0.025,0.005),(0.05,0.01)\}~$p.u.	\label{fig:test3.kp.kd}}
\end{figure}

Fig.~\ref{fig:test3.kp.kd} shows impact of dual-port GFM control gains $(k_p,k_d)$ on the magnitude of transfer functions from the load step $ P_L$ to the VSC frequency $G_{P_\text{L},\omega_\vsc}$ (left) and DC voltage $G_{P_\text{L},v_\dc}$ (right). 

\begin{figure*}[t]
	\centering 
	\includegraphics[width=1\textwidth, trim={0 1cm 0 0}, clip]{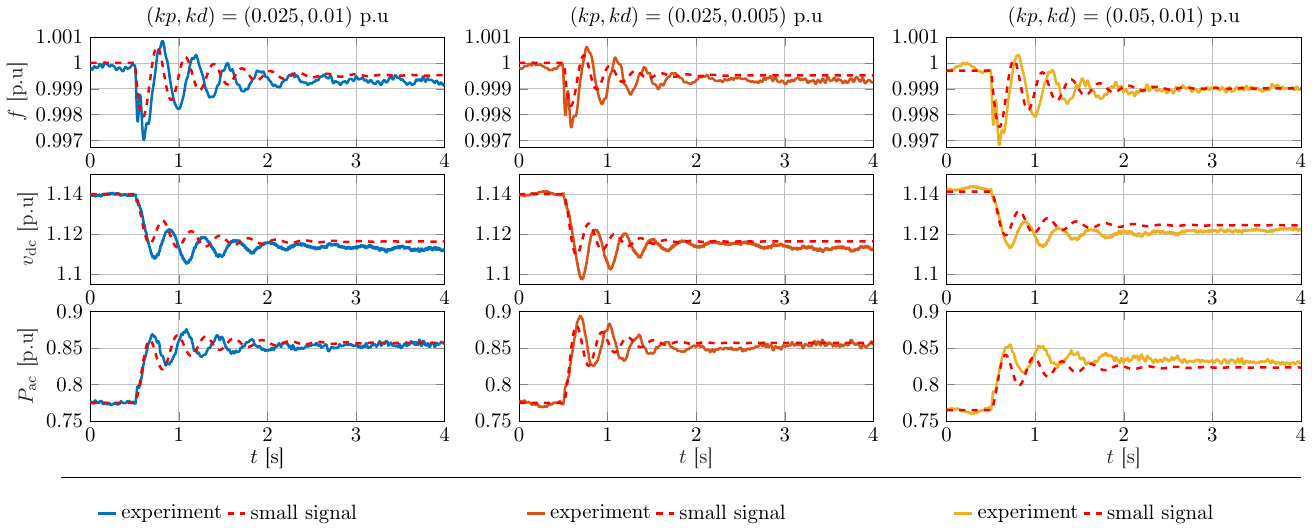} 
	\caption{Comparison of the small-signal model (dashed) and experimental results (solid) for the system in Fig.~\ref{fig:simplified.setup}~(a) and a a load step of $2.5$~kW. \label{fig:test3.model.validation}}
\end{figure*}

\begin{figure*}[t!]
\centering
  \includegraphics[width=1\textwidth]{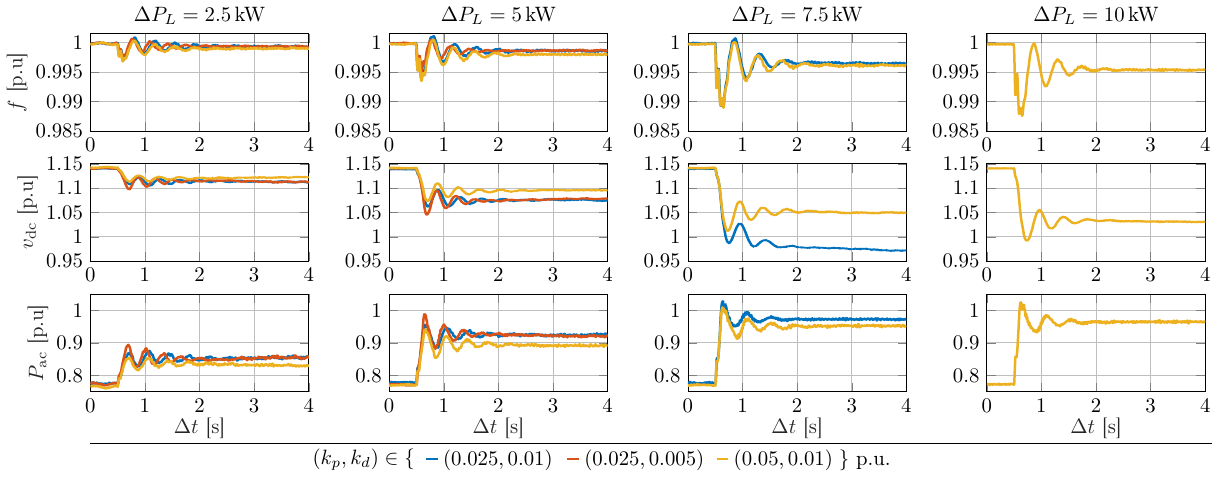}
  \caption{Experimental results of the test case in Fig.~\ref{fig:simplified.setup}~(a) for different control gains and load steps $\Delta P_L \in \{2.5,5,7.5,10\}$~kW. \label{fig:test3.measurements}}
\end{figure*}
As expected based on the open-loop controller transfer function, the proportional gain $k_p$ only significantly influences the steady-state while the derivative gain $k_d$ has significant impact on the transient response. In particular, decreasing the derivative gain $k_d$ shifts the resonant frequency peak towards higher frequencies. While the impact of the derivative gain on the resonant peak is largely negligible for $G_{P_\text{L},v_\dc}$, decreasing $k_d$ lowers the gain of the resonant peak for $G_{P_\text{L},\omega_\vsc}$. Moreover, we note that the derivative gain $k_d$ can be used to compensate the destabilizing effects of line dynamics~\cite{G22}. In contrast, while increasing $k_p$ slightly reduces the magnitude of the resonant peak, the impact of $k_p$ on oscillation damping and the transient response is not significant.

\subsection{Experimental results}
Next, we illustrate and validate the small-signal model and findings using the experimental testbed. To this end, we assume a nominal load of $P_L=20$~kW, $P_\sg^\star \approx 7$~kW, and $P_\pv^\star \approx 13$~kW. Fig.~\ref{fig:test3.model.validation} shows the response of the VSC frequency, DC voltage, and VSC active power to a load step of $\Delta P_L=2.5$~kW for the testbed (solid-line) and small-signal model (dashed-line). It can be seen that the small-signal model and experimental results are largely in agreement both with respect to the steady-state and transient response.

Next, load variation $\Delta P_L \in \{2.5, 5, 7.5,10\}$~kW are used to perturb the system and illustrate the response of the VSC under different control gains. Fig.~\ref{fig:test3.measurements} shows experimental results illustrating that frequency of transient oscillations are approximately the same for identical $k_d$ (blue and yellow signals) and the damping of transient oscillation is higher for larger $k_p$ leading to smaller settling time. Moreover, reducing the derivative gain $k_d$ leads to a larger initial drop in the DC voltage which can have a negative impact and trigger the DC voltage protection during large disturbances. This was observed for  $\Delta P_L=\{7.5,10\}$~kW. Hence, results for $k_d=0.005$~p.u. (red) are not shown in Fig.~\ref{fig:test3.measurements} for $\Delta P_L=\{7.5,10\}$~kW. Similarly, decreasing $k_p$ leads to increased DC voltage droop gain (i.e., $ v_\dc =  \omega_\vsc / k_p$) which can trigger DC voltage protection during large disturbances. This was observed for $\Delta P_L =10$~kW. Hence, results for $k_p=0.05$ are not shown in Fig.~\ref{fig:test3.measurements} for $\Delta P_L =10$~kW. The load step of $7.5$~kW is relatively large ($\approx 41.2$\% of the PVs rated power) and, hence, taking operating limits into account in the control tuning is required.

\section{LVDC Connection of an LVAC system with SG to a Utility grid} \label{sec:LVDCasync}
In this section, we consider connecting a LVAC system (i.e., AC~1 with SG and load) to the public utility grid solely via LVDC (see Fig.~\ref{fig:simplified.setup}~(b)). Conceptually, the system topology is similar to HVDC connecting two asynchronous areas~\cite{GSP+21,SG23ifac} but voltage and the power rating significantly differ. Because the power rating of the utility grid is much larger than that of AC~1, the grid can be approximated as an infinite bus (i.e. SG with infinite inertia). Consequently, load steps in AC~1 do not lead to steady-state frequency changes in AC~2. Hence, we use this configuration to characterize and validate the transient response of dual-port GFM control. First, we simplify the general small-signal model developed in Sec.~\ref{sec:general.small.signal.model} to reflect the testbed configuration considered in this section. Next, we analyze the transient response. Finally, we validate the model and findings against the hardware testbed and provide additional experimental results. 

\subsection{Small-signal model}
\begin{figure}[b]
	\centering 
	\includegraphics[trim={1.95cm 0.3cm 1.95cm 0.4cm},clip, scale=0.825]{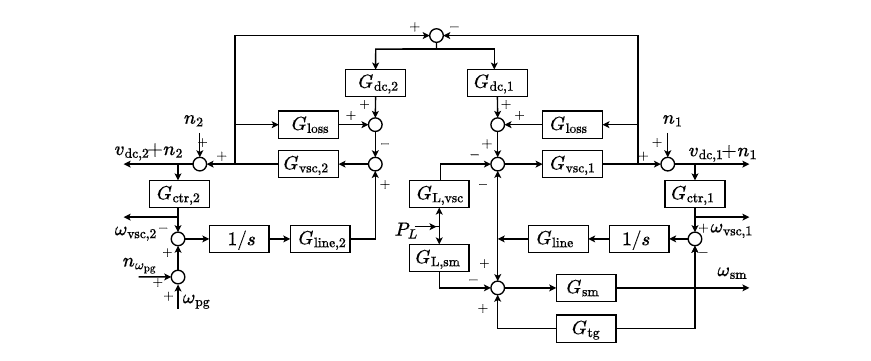} 
	\caption{Block diagram of the testbed configuration depicted in Fig.~\ref{fig:simplified.setup}~(b).	\label{fig:test7_blk_diag}}
\end{figure}

The system in Fig.~\ref{fig:simplified.setup}~(b) consists of a SG (with $k_\tg=20$) and load (AC~1), the connection to the public utility grid (AC~2), and an LVDC connection between AC~1 and AC~2. Modeling the utility grid as an input with variable frequency $\omega_\pg$ results in the block diagram is shown in Fig.~\ref{fig:test7_blk_diag}. 
Because the main focus of the manuscript is frequency dynamics, the AC voltage magnitude $V_\ac^\star$ of the utility grid is assumed to be nominal. While this assumption is not satisfied in general, it suffices to reason about frequency dynamics in our setting. 

The transfer functions corresponding to the LVDC network power and losses are given by $G_{\dc,i}=v_{\dc,i}^\star/(\ell_\dc s+r_\dc)$ for $i\in \{1,2\}$, $G_{\text{loss}} = (v_{\dc,2}^\star-v_{\dc,1}^\star) / (\ell_\dc s+r_\dc)$.  The transfer function of the connection to the public grid is given by $G_{\text{line},2} ={V_\ac^\star}^2 \omega^\star k_{\ell,2}\ell_{\pg}/(s^2+2\rho_2 s + \rho_2^2 + (\omega^\star k_{\ell,2})^2)$, where $\rho_{2}=r_\pg/\ell_\pg$ and $k_{\ell,2}=(\tilde{\ell}+\ell_\pg)/\ell_\pg$. The remaining transfer functions are the same as in Sec.\ref{sec:PV.small.signal.model}. Finally, note that the closed-loop transfer functions depends on the linearization points $v_{\dc,1}^\star$ and $v_{\dc,2}^\star$ (i.e., if $v_{\dc,1}^\star=v_{\dc,2}^\star$, then $G_{\text{loss}}=0$).

\subsection{Baseline VSC control tuning}
We first use analytic steady-state and stability results to obtain an initial set of control gains.

\subsubsection{Steady-state} Unless stated otherwise, we chose the control gains $k_{p,1}=k_{p,2}=0.025$~p.u., i.e., a $10$\%  VSC DC bus voltage deviation maps to a frequency deviation of $0.25$\% (i.e., $125$~mHz in a $50$~Hz system). This choice is motivated by the aim to make the frequency response of VSCs compatible with the stiff frequency dynamics of the infinite bus by tightly controlling the frequency at the VSC terminal. Moreover, $k_{d,1}=k_{d,2}=0.001$~p.u. are used to induce damping in the DC voltage dynamics.

\subsubsection{Stability} When neglecting the dynamics of AC and DC connections, stability of the system dynamics can be guaranteed using the conditions developed in~\cite{SG24,SG23ifac} for quasi-steady-state network models. However, note that depending on, e.g., the inductance and $X/R$ ratio of the AC and DC networks the analytical conditions in~\cite{SG24,SG23ifac} may be misleading. Therefore, we only use \cite[Thm.~3]{SG23ifac} to obtain a baseline set of parameters and will investigate the impact of the control gains in the presence of line dynamics numerically. To this end, note that the system in Fig.~\ref{fig:simplified.setup}~(b) contains an SG (with $k_\tg=20>0$). Hence,~\cite[Cond.~1]{SG23ifac} is trivially satisfied. Moreover, to ensure stability, the dual-port GFM control gains need to satisfy~\cite[Cond.~2]{SG23ifac}. For the system under study, this results in $k_{d,1}/k_{p,1}<0.2571$ and $k_{d,2}/k_{p,2}<0.5142$. For $k_{p,1}=k_{p,2} = 0.025$~p.u., this results in the bounds $k_{d,1}<0.0064$~p.u. and $k_{d,2}< 0.0129$~p.u. on the derivative control gains. We observe that~\cite[Cond.~2]{SG23ifac} is conservative and, in practice, larger derivative gains $k_{d,1}$ and $k_{d,2}$ can be selected. Next, we use~\cite[Cond.~2]{SG23ifac} as a starting point and leverage the small-signal model shown in Fig.~\ref{fig:test7_blk_diag} to tune the derivative gains $k_d$.

\subsection{Transient response}\label{subsec:LVDC:transient}
\begin{figure}[t]
	\centering 
	\includegraphics[width=\columnwidth]{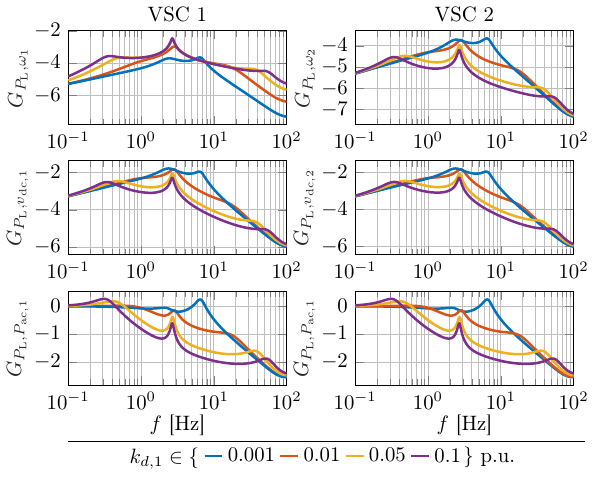} 
	\caption{Bode magnitude plots of the transfer functions from the load $P_L$ to the VSC frequency, DC voltage, and power injection for $k_{p,1}=k_{p,2}=0.025$~p.u., $k_{d,2}=0.001$~p.u. and the model shown in Fig.~\ref{fig:test7_blk_diag}.	\label{fig:test7_bode}}
\end{figure}

The small-signal model is used to characterize the dynamic response of the system. 
Figure~\ref{fig:test7_bode} shows the Bode magnitude plot of the closed-loop transfer functions $G_{P_\text{L},\omega_{\vsc,i}}$, $G_{ P_\text{L},v_{\dc,i}}$, and $G_{ P_\text{L},P_{\ac,i}}$ from the load $P_L $ to the VSC frequency, DC voltage, and power injection for $i\in \{1,2\}$. The dual-port GFM control gains are chosen as $k_{p,1}=k_{p,2}=0.025$~p.u., $k_{d,2}=0.001$~p.u., and  $k_{d,1} \in \{ 0.01,0.001,0.05,0.1\}$~p.u. is used to illustrate the impact of the derivative gain.
It can be observed that increasing $k_{d,1}$ shifts the resonant frequencies of all transfer functions shown in Fig.~\ref{fig:test7_bode} to lower frequencies. Notably, for larger $k_{d,1}$ resonance peaks are present at a very low frequency range ($\approx 0.2\,$Hz). Moreover, increasing $k_{d,1}$ lowers the resonance peaks in the transfer functions from the load disturbance $P_L$ to the VSC AC power injection $P_\ac$ and DC voltage $v_\dc$. In contrast, the impact on the transfer function from the load disturbance $P_L$ to the VSC frequency $\omega$ is more nuanced. Specifically, increasing $k_{d,1}$ increases the gain of $G_{P_L,\omega_1}$ at all frequencies (including the resonance frequency), while it decreases the gain at the resonance frequency for $G_{P_L,\omega_2}$. In other words, increasing $k_{d,1}$ increases the damping of the dynamics of $v_{\dc,1}$ at the expense of reduced damping of $\omega_1$. In contrast, because the load $P_L$ is in AC~1, the load disturbance can only impact $\omega_2$ through the LVDC connection. Thus, increased damping of $v_{\dc,1}$ results in increased damping of $v_{\dc,2}$ and $\omega_2$.

\subsection{Experimental results}
\begin{figure}[b!!!]
	\centering 
	\includegraphics[width=1\columnwidth]{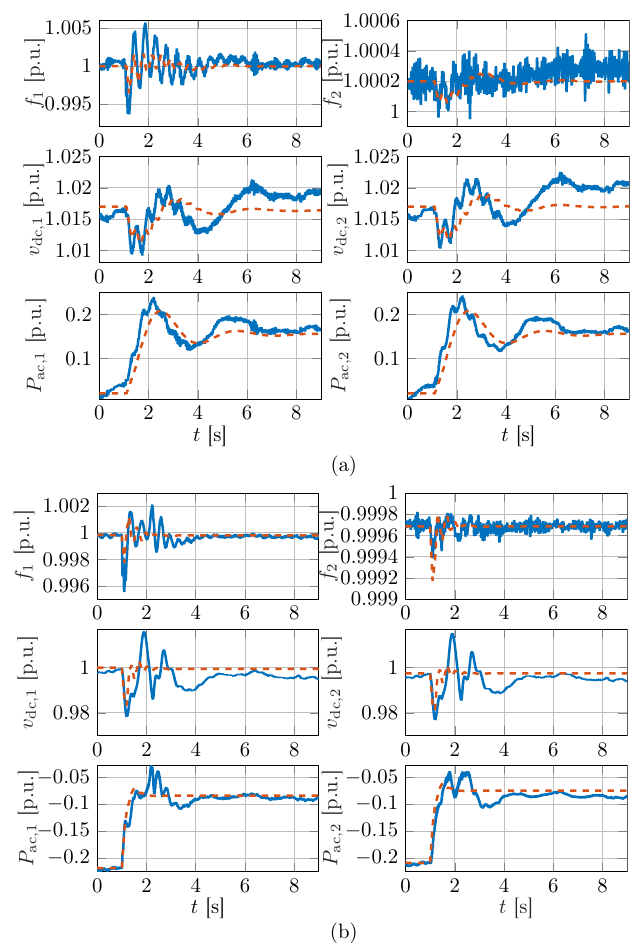} 
	\caption{
    Comparison of the small-signal model (dashed) and measurements (solid) for a $3$~kW load step for the system in Fig.~\ref{fig:simplified.setup}~(b) and $k_{p,1}=k_{p,2} =0.025$~p.u., $k_{d,2}=0.001$~p.u., (a) $k_{d,1} = 0.01$~p.u., $v_{\dc,1}^\star -v_{\dc,2}^\star = 0$~p.u. and (b) $k_{d,1} = 0.1$~p.u., $v_{\dc,1}^\star -v_{\dc,2}^\star = 0.0018$~p.u.\label{fig:test7_validation1}}
\end{figure}

To validate the small-signal model we apply a load step of $3$~kW to the experimental testbed. The measured (solid-line) and small-signal model (dashed-line) responses of the VSC frequency, DC voltage and power for different control parameters are shown in  Fig.~\ref{fig:test7_validation1}. 
While the general trends predicted by the small signal model match the measurements, the frequency and magnitude of oscillations do not perfectly coincide. The main reason for this mismatch are modeling simplifications such as modeling the utility grid as an infinite bus, neglecting AC voltage dynamics at the point of connection to the utility grid, and estimates used to parametrize the grid impedance. Nevertheless, the small-signal model is still useful to analyze the dynamic response of the system. Finally, we note that during the experiment the utility grid frequency deviates from its nominal point, resulting in a significant difference between the steady-state of the experimental results and small-signal model (see Fig.~\ref{fig:test7_validation1}~(a)). As expected, dual-port GFM control adapts to this steady-state frequency deviation. In other words, the steady-state DC voltage deviation is proportional to the frequency deviation according to the gain $k_{p,i}$.

\begin{figure}[b!]
	\centering 
	\includegraphics[width=1\columnwidth]{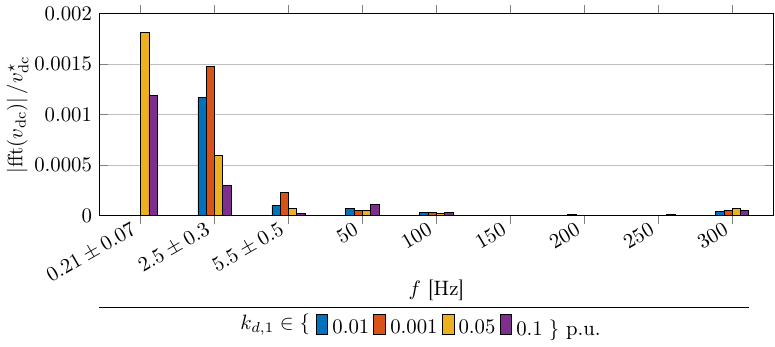} 
	\caption{Frequency spectrum of the DC voltage of VSC~1 for $k_{p,1}=k_{p,2}=0.025$~p.u., $k_{d,2}=0.001$~p.u. and $k_{d,1}= \{ 0.01,0.001,0.05,0.1\}$~p.u. and the load step of $3$~kW. \label{fig:test7_histo}}
\end{figure}

In addition, to validate the analysis in Sec.~\ref{subsec:LVDC:transient} the FFT of the DC voltage measurements for different $k_{d,1}= \{ 0.01,0.001,0.05,0.1\}$~p.u. and a load step of $3$~kW is illustrated in Fig.~\ref{fig:test7_histo}. As predicted by the analytical model, increasing the gain $k_{d,1}$ moves the resonance peak to lower frequencies and lowers its magnitude.

\begin{figure}[b!]
	\centering 
	\includegraphics[width=\columnwidth]{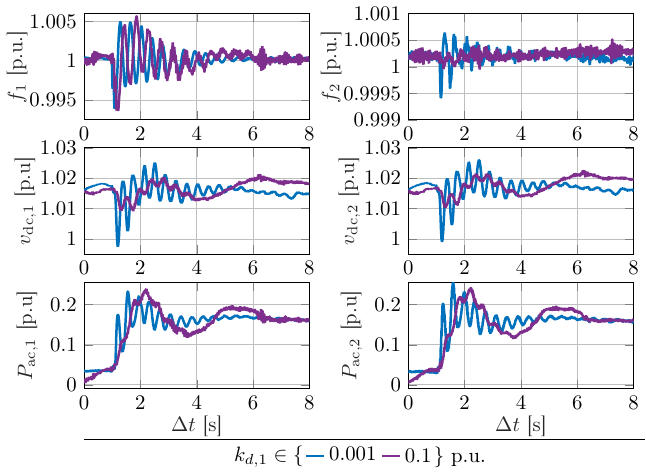} 
	\caption{Experimental results for the system in Fig.~\ref{fig:simplified.setup}~(b), $k_{d,2}=0.001$~p.u., $k_{d,1}\in \{0.001,0.1\}$~p.u., $k_{p,1}=k_{p,2}=0.025$~p.u. and a $3$~kW load step. \label{fig:test7.diff.kd}}
\end{figure}

Figure~\ref{fig:test7.diff.kd} shows measurements for two values of the control gains $k_{d,1}\in \{0.001,0.01\}$~p.u. and a $3$~kW load step. 
We again observe that increasing $k_{d,1}$ shifts the transient oscillations of $v_\dc$, $P_\dc$ and $\omega_{\vsc,2}$ to lower frequencies and increases the damping, i.e., oscillations in AC~1 are not mapped to AC~2.  Finally we note that while higher $k_{d,1}$, provides less damped $\omega_{\vsc,1}$ response, this impact is minimal and it does not deteriorate system's behavior. 

\begin{figure}[t!]
	\centering 
	\includegraphics[scale=0.84]{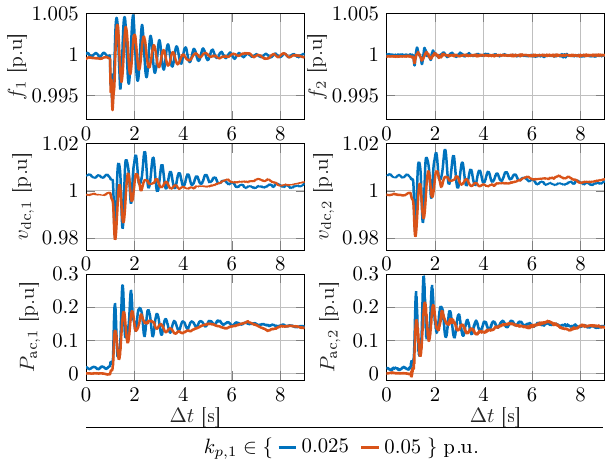} 
	\caption{Experimental results for the system Fig.~\ref{fig:simplified.setup}~(b),  $k_{p,2}=0.025$~p.u., $k_{d,1} = k_{d,2} = 0.001$~p.u., $k_{p,1} \in \{0.025, 0.05\}$~p.u., and a $3$~kW load step. \label{fig:test7.diff.kp}}
\end{figure}

Finally, Fig.~\ref{fig:test7.diff.kp} shows measurements for two values of the proportional gains $k_{p,1} \in \{0.025, 0.05\}$~p.u. for a $3$~kW load step, illustrating that increasing $k_{p,1}$ does not change the frequency of the transient oscillations but provides increased damping. However, because $k_d=0.001$ is relatively small, frequency oscillations in AC~1 are propagated to AC~2.

\section{Parallel LVDC and LVAC connection} \label{sec:LVDCsync}
Finally, we experimentally validate connecting an LVAC system to the utility grid via LVDC and LVAC as shown in Fig.~\ref{fig:simplified.setup}~(c). In particular, we focus on dispatching the power flow through the parallel LVAC and LVDC connections when using dual-port GFM control. Moreover, we investigate the response to load step in the LVAC system.

Notably, dual-port GFM control does not explicitly control the converter AC or DC power injection. Therefore, the nominal power flow through the converter and LVDC link can only be dispatched via the converter DC voltage setpoints $v_\dc^\star$~\cite[Sec.~V-B]{GSP+21}. To demonstrate and validate the ability to control the LVDC power flow, the experimental testbed is configured as shown in Fig.~\ref{fig:simplified.setup}~(c) and  the converter control gains are chosen as $k_{p,1}=k_{p,2}=0.025$~p.u. and $k_{d,1}=k_{d,2}=0.001$~p.u., i.e., in steady state a $10$\% VSC DC bus voltage deviation maps to a $0.25$\% frequency deviation. Due to the LVAC connection, the two AC systems AC~1 and AC~2 are synchronous, i.e., $f_1=f_2$ in steady state.

\begin{figure}[t!]
	\centering 
	\includegraphics[width=\columnwidth]{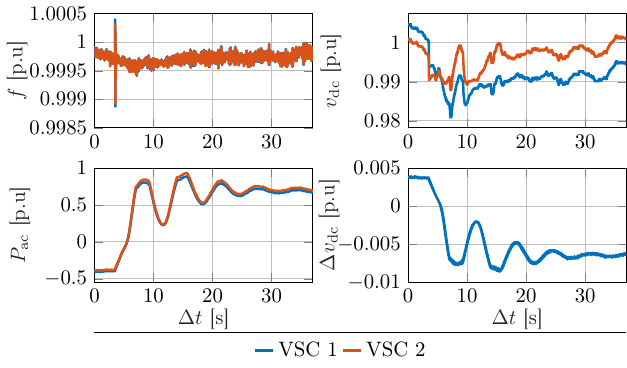} 
	\caption{Dispatch of the LVDC power flow for the system in Fig.~\ref{fig:simplified.setup}~(c). The DC voltage setpoints are changed from $(v_{\dc,1}^\star,v_{\dc,2}^\star) = (1.0037,1)$~p.u. to $(v_{\dc,1}^\star,v_{\dc,2}^\star) = (0.9975,1)$~p.u. at $t\approx 3.5\,$s and $\Delta v_\dc \coloneqq v_{\dc,1}-v_{\dc,2}$. \label{fig:test9_1}}
\end{figure}

Initially, power is flowing from the utility grid to the load $P_L$ through the LVAC connection and power is circulated back to the utility grid through the LVDC connection (i.e., $v_{\dc,1}^\star>v_{\dc,2}^\star$). In particular, $(v_{\dc,1}^\star,v_{\dc,2}^\star) = (1.0037,1)$~p.u. is chosen as initial operating point. Since circulating power through the LVDC and LVAC connections is undesirable, the DC voltage setpoints are changed to $(v_{\dc,1}^\star,v_{\dc,2}^\star) = (0.9975,1)$~p.u. at $t\approx 3.5\,$s to redispatch the power flow such that power is flowing to the load both through the LVAC and LVDC connection. Experimental results are shown in Fig.~\ref{fig:test9_1}.
It can be seen that the DC voltage difference $v_{\dc,1}-v_{\dc,2}$ on (bottom left) changes sign in response to the setpoint update and, consequently, the AC power injection of VSC~1 (bottom right) changes direction verifying that the nominal steady-state power flow across the LVDC link can be controlled through DC voltage setpoints. At the same time, because the controller does not explicitly control power, the dynamics of the LVDC line (i.e., cable inductance and capacitance) have significant impact on transient behavior. Similar behavior (i.e., slower transient response, but without transient oscillations) is also observed in~\cite[Fig.~21]{GSP+21}. Note that~\cite{GSP+21} examines HVDC link, while in this manuscript considers an LVDC connection with significantly different line and system parameters. Investigating this aspect in detail and developing active LVDC network  damping controls to improve the response to dispatch signals is seen as interesting topic for future work.

\begin{figure}[b!]
	\centering 
	\includegraphics[width=\columnwidth]{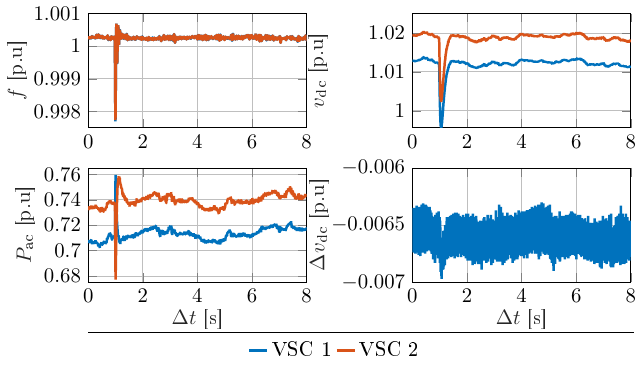} 
	\caption{Experimental results for a $5$~kW step increase of the load $P_L$ at $t\approx 1\,$s and the system configuration shown in Fig.~\ref{fig:simplified.setup}~(c). \label{fig:test9_2}}
\end{figure}

Finally, Fig.~\ref{fig:test9_2} illustrates the response of the VSCs and LVDC connection during a $5$~kW step increase of the load $P_L$. For this experiment, the control gains $k_{p,1} =k_{p,2}=0.025$~p.u. and $k_{d,1} = k_{d,2} = 0.005$~p.u. and DC voltage setpoints $(v_{\dc,1}^\star,v_{\dc,2}^\star) = (0.9938,1)$~p.u. have been used. 
It can be observed that, apart from a large but brief initial transient, the power flow across the LVDC link is maintained and the load is supplied through the LVAC line. This result can be explained by the utility grid approximately corresponding to an infinite bus. In other words, the utility grid frequency is insensitive to the load change. At the same time, AC~1 and AC~2 are synchronously connected through the LVAC connection and, in steady state, the frequency is near nominal through the system. Therefore, due to the frequency synchronization, the power flow through the VSCs and LVDC connection will remain approximately at the (nominal) pre-event values for both dual-port GFM control and $P_\ac-f$ droop control.

It should also be noted that the LVDC and LVAC lines are connected to the same bus and that the losses of the LVDC in the experimental testbed are significantly higher than the losses of the LVAC line (approximately one order of magnitude). In other words, supplying the increased load through the LVAC connection is preferable in this scenario.

\section{Conclusion} \label{sec:conclusion}
This work presented and validated a general small-signal dynamical model for hybrid AC/DC power systems. In addition, the study provided a quantitative characterization of the dynamic behavior of dual-port grid-forming (GFM) control. Specifically, it is shown that the derivative gain primarily influences the frequency of closed-loop transient oscillations, while the proportional gain determines the steady-state droop and contributes to additional damping during transients. Furthermore, the derivative gain is demonstrated to be effective in compensating for transmission line dynamics, thereby enabling systematic shaping of the system’s transient response. Finally, the manuscript analyzes and experimentally demonstrates the application of dual-port GFM control to low-voltage (LV) DC networks and hybrid LVDC/LVAC networks. This setting is particularly relevant for DC microgrids deployed at the distribution level or within local residential clusters, where households are equipped with distributed renewable energy sources. While the initial results are promising, future work should investigate the impact of converter constraints such as DC voltage limits, current limits, and unbalanced AC grid conditions on control and dynamics of hybrid LVAC/LVDC systems.

%
\bibliographystyle{IEEEtran}
\bibliography{IEEEabrv,bib_file}

\newpage
\newpage

\end{document}